\documentclass[manuscript]{acmart}
\AtBeginDocument{%
  }

\usepackage{todonotes}
\usepackage{xspace}
\usepackage{booktabs,tabularx,array}
\usepackage{listings}
\usepackage{enumitem}
\usepackage{svg}
\usepackage{soul}
\usepackage{fancyref}

\newcounter{theme}
\renewcommand{\thetheme}{T\arabic{theme}}
\newcommand{\themelabel}[1]{\refstepcounter{theme}\label{#1}\thetheme}

\newcommand{\PaperWordCount}{%
  \immediate\write18{texcount -1 -sum -merge -q "paper.tex" > "\jobname.wc"}%
  \textcolor{red}{\input{\jobname.wc}words}%
}

\setcopyright{acmlicensed}
\copyrightyear{2025}
\acmYear{2025}
\acmDOI{XXXXXXX.XXXXXXX}

\begin{document}

\title{\retrace: Interactive Visualizations for Reasoning Traces of Large Reasoning Models}

\author{Ludwig Felder}
\affiliation{%
  \institution{Technical University of Munich}
  \city{Heilbronn}
  \country{Germany}}
\email{ludwig.felder@tum.de}

\author{Jacob Miller}
\affiliation{%
  \institution{Technical University of Munich}
  \city{Heilbronn}
  \country{Germany}}
\email{jacob.miller@tum.de}

\author{Markus Wallinger}
\affiliation{%
  \institution{Technical University of Munich}
  \city{Heilbronn}
  \country{Germany}}
\email{markus.wallinger@tum.de}

\author{Stephen Kobourov}
\affiliation{%
  \institution{Technical University of Munich}
  \city{Heilbronn}
  \country{Germany}}
\email{stephen.kobourov@tum.de}

\author{Chunyang Chen}
\affiliation{%
  \institution{Technical University of Munich}
  \city{Heilbronn}
  \country{Germany}}
\email{chun-yang.chen@tum.de}
\renewcommand{\shortauthors}{Felder et al.}
\newcommand{\retrace}{\textsc{Re}\textsc{Trace}\xspace}
\begin{abstract}

Recent advances in Large Language Models have led to Large Reasoning Models, which produce step-by-step reasoning traces. These traces offer insight into how models think and their goals, improving explainability and helping users follow the logic, learn the process, and even debug errors. These traces, however, are often verbose and complex, making them cognitively demanding to comprehend. We address this challenge with \retrace, an interactive system that structures and visualizes textual reasoning traces to support understanding. We use a validated reasoning taxonomy to produce structured reasoning data and investigate two types of interactive visualizations thereof. In a controlled user study, both visualizations enabled users to comprehend the model’s reasoning more accurately and with less perceived effort than a raw text baseline. The results of this study could have design implications for making long and complex machine-generated reasoning processes more usable and transparent, an important step in AI explainability.

\end{abstract}



\begin{teaserfigure}
    \centering
    \includegraphics[width=\linewidth]{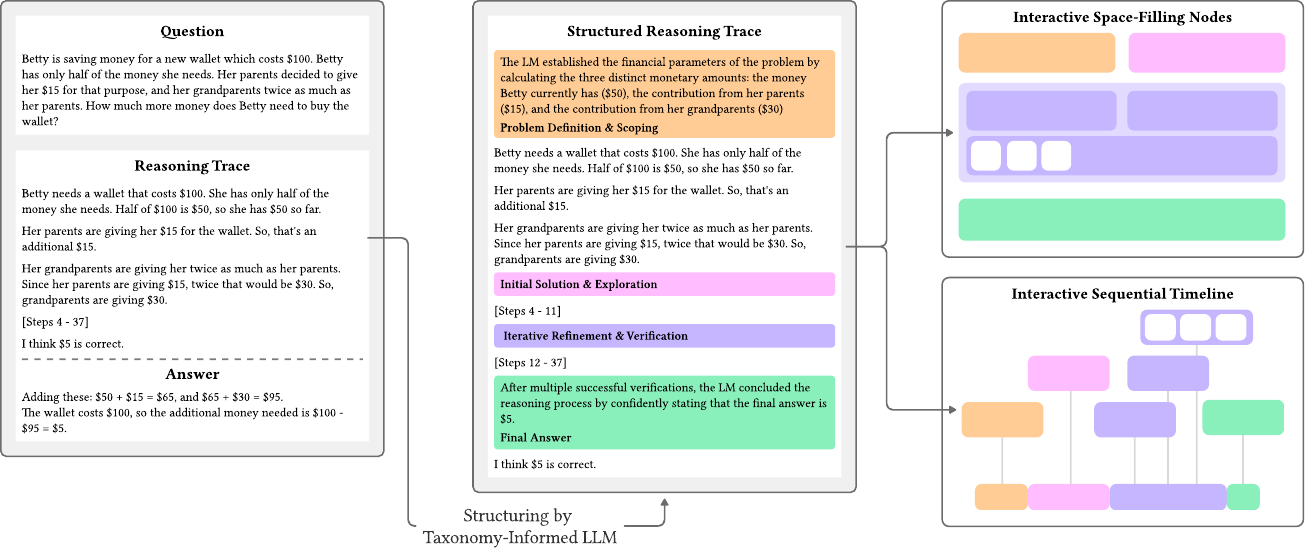}
    \caption{\retrace lets users explore a Large Reasoning Model's reasoning with interactive visualizations. \retrace structures the raw textual trace (left) using an external LLM guided by a reasoning taxonomy, generates a labeled and summarized trace (center), and renders it as two distinct interactive visualizations: Space-Filling Nodes and Sequential Timeline (right).
    }
    \Description{Diagram of the ReTrace system's three-stage process. The diagram shows a text box of a raw "Reasoning Trace" on the left. An arrow points to a central box containing a "Structured Reasoning Trace," which has been grouped, summarized and labeled. A final arrow points from this structured trace to two distinct visualizations on the right: one hierarchical layout called "Interactive Space-Filling Nodes" and one linear layout called "Interactive Sequential Timeline".}
    \label{fig:enter-label}
\end{teaserfigure}
\maketitle

\section{Introduction}
Large Language Models (LLMs) have recently found success in shifting from ``answer directly” to ``reason, then answer.” The discovery of Chain-of-Thought (CoT) prompting has shown that adding a series of step-by-step reasoning examples to a given prompt significantly improves the LLM's ability to perform complex reasoning~\cite{reynoldsPromptProgrammingLarge2021, kojimaLargeLanguageModels2023, weiChainofThoughtPromptingElicits2023a}. For example, if a user were to ask for the answer to a mathematical problem, they would add several step-by-step solutions of other math problems. While effective, this method of manually constructing reasoning examples makes the already challenging prompting process~\cite{zamfirescu-pereiraWhyJohnnyCant2023a} even more demanding for users. In contrast to traditional LLMs, Large Reasoning Models (LRMs)~\cite{Claude3S, deepseek-aiDeepSeekR1IncentivizingReasoning2025a, openaiOpenAIO1System2024, yangQwen3TechnicalReport2025} are explicitly trained to perform this multi-step CoT natively, thereby generating a reasoning trace as a computational by-product to the final answer~\cite{muennighoffS1SimpleTesttime2025a, kumarLLMPostTrainingDeep2025}. In other words, the model ``thinks out loud'' and externalizes part of its computation process in natural language~\cite{korbakChainThoughtMonitorability2025}.

These traces matter. As a readable artifact, they expose decision points, verification moves, and failure modes that are otherwise hidden when relying only on the final answer. For humans, reasoning traces create an opportunity to investigate how an LRM generated an answer rather than only what the answer is. Although prior work has not evaluated LRM reasoning traces directly, adjacent work in explanatory debugging and human-grounded explainable AI (XAI) shows that exposing intermediate, rationale-like artifacts can improve mental models and help people surface errors \cite{doshivelezRigorousScienceInterpretable2017, kuleszaPrinciplesExplanatoryDebugging2015}. Studies that evaluate explanations through forward and counterfactual simulation suggest that such visibility can help people anticipate model behavior, which is central when deciding to rely on a result \cite{haseEvaluatingExplainableAI2020}. Our work, therefore, focuses on making traces analyzable, surfacing their inner structure while preserving provenance as well as in supporting sensemaking. 

In practice, raw traces are rarely readable at scale. They are often very long and verbose, with LRMs like DeepSeek-R1-Distill-Qwen-1.5B generating on average over 15,000 tokens on math-related questions~\cite{hou2025thinkprune}. Analyses of reasoning traces reveal overthinking patterns such as excessive planning with little progress, abandoning tasks too early, and issuing multiple actions in a single iteration without waiting for feedback, which further increases their complexity~\cite{cuadronDangerOverthinkingExamining2025}. LRMs are inherently encouraged to ``think long'' \cite{muennighoffS1SimpleTesttime2025a,kumarLLMPostTrainingDeep2025}. These large volumes of generated text hinder sensemaking \cite{jiangGraphologueExploringLarge2023a,geroSupportingSensemakingLarge2024}. Simple formatting or summarization helps with skimming but collapses provenance, undermining the details-on-demand needed for careful inspection and learning \cite{shneiderman2003eyes}. Our work addresses this gap by improving the readability of these reasoning traces.

One way to address the complexity of large token sequences is through visualization. By making reasoning processes more explainable, visualization helps users discern structure, generate insights, and communicate findings effectively \cite{KucherK15}. Prior work observed that reasoning traces have an identifiable internal structure~\cite{marjanovicDeepSeekR1ThoughtologyLets2025, hammoudLastAnswerYour2025}. Coherent phases, such as rephrasing, decomposition, candidate generation, and evaluation, into which the reasoning trace can be broken down. Visualizations can leverage this structure to visually summarize while maintaining links back to the raw thoughts. 
Interactive visualizations, in particular, can support established sensemaking strategies by providing high-level overviews while preserving access to low-level details, like reasoning steps, on demand \cite{heerInteractiveDynamicsVisual2012, shneiderman2003eyes, pirolliSensemakingProcessLeverage}.

We present \retrace, an interactive system that structures and visualizes LRM traces to support human sensemaking. In an exploratory study with early prototypes and four experts, we surfaced key design challenges. From this, we distilled three design goals: make the reasoning sequence explicit, offload cognitive work via multi-level summaries, and enable detail on demand. \retrace utilizes an LLM-driven pipeline, grounded in a validated reasoning taxonomy~\cite{marjanovicDeepSeekR1ThoughtologyLets2025}, to transform unstructured, verbose reasoning traces into a structured, interactive visualization. Our interface renders this data into two distinct interactive visualizations: a hierarchical \textit{Space-Filling} layout that surfaces strategic structure and a \textit{Sequential Timeline} that preserves chronological flow. By linking a high-level overview to detailed access to the verbatim reasoning ``thoughts'' of LRM on demand, our design aims to reduce cognitive load and make the model's strategy more apparent to users.

To evaluate the utility of \retrace, we conducted a within-subjects user study with 18 participants from a wide variety of backgrounds, including non-computing backgrounds. We compared our two interactive visualizations against a raw text baseline, measuring their impact on user comprehension, perceived workload, and confidence in judging the model's process. Our findings reveal that the structured visualizations provided by \retrace improve users' ability to summarize the model's strategy and reduce cognitive load compared to the raw trace baseline.

Our contributions include:
\begin{itemize}
\item A two-level visualization approach for LRM reasoning traces that preserves chronological sequence while revealing high-level strategy through semantic grouping.
\item The design and implementation of \retrace, an interactive visualization system that parses, summarizes, and visualizes reasoning traces with progressive disclosure to support efficient overview–detail analysis. 
\item Findings from a user study with 18 participants that reveal the utility of \textsc{ReTrace} in improving comprehension and reducing perceived workload relative to raw traces. 
\end{itemize}

\section{Background and Related Work}
In this section, we first give some background on language model reasoning followed by established HCI principles for visualization, abstraction, and interaction techniques. We then end the section with a review of related work to situate our contribution in context of the state-of-the-art. 

\subsection{Reasoning in Language Models}
The pursuit of computational reasoning has been a long-standing goal in natural language processing~\cite{yuNaturalLanguageReasoning2023}, and has seen a recent surge with the advent of LLMs~\cite{huangReasoningLargeLanguage2023}.  A pivotal breakthrough was Chain-of-Thought (CoT) prompting, which demonstrated that eliciting a step-by-step rationale before the final answer significantly improves performance on complex reasoning tasks~\cite{weiChainofThoughtPromptingElicits2023a}. This technique revealed that LLMs possess emergent reasoning capabilities that can be triggered with simple cues, effectively making them zero-shot reasoners~\cite{kojimaLargeLanguageModels2023}. The success of these trace-based methods motivated a shift from in-context prompting to explicitly fine-tuning models for reasoning. Early approaches involved bootstrapping, where models use their own self-generated rationales as training signals~\cite{zelikmanSTaRBootstrappingReasoning2022}. 
These efforts culminated in the development of dedicated Language Reasoning Models (LRMs)~\cite{Claude3S, deepseek-aiDeepSeekR1IncentivizingReasoning2025a, openaiOpenAIO1System2024, yangQwen3TechnicalReport2025}. As a result, LRMs are designed to produce a coherent, multi-step reasoning trace as a distinct, pre-answer step, making the trace itself a first-class output of the model.

The high-level perspective provided by the reasoning trace positions it as a critical object for monitoring and auditing. As~\citet{korbakChainThoughtMonitorability2025} highlights, treating a CoTs as observable ``latent variables in the model's computation'' presents a significant opportunity for AI safety and alignment. Our work, therefore, focuses on the sensemaking challenges posed by the high-level, pre-answer reasoning traces from LRMs. We leverage the internal structure identified in these traces to design an interactive visualization system that supports human comprehension and auditing of the model's articulated reasoning process.

\subsection{Sensemaking of Text Data} 
A primary challenge in comprehending LRM outputs stems from their extreme verbosity. 
Prior work shows that large amounts of text negatively impacts user's ability to understand and interact with an output~\cite{jiangGraphologueExploringLarge2023a}. As eye-tracking studies on reading demonstrate, rather than sequential reading, users tend to skim through text to cope with the volume of available information~\cite{duggan2011skim}. ~\citet{raynerMuchReadLittle} observe that reading can be contrasted with skimming, by which one's eyes move quickly through the text to get a general idea of the content. While efficient, skimming requires focused attention and presents its own cognitive challenges ~\cite{duggan2011skim}, creating a clear need for interfaces that provide explicit skimming support~\cite{geroSupportingSensemakingLarge2024}. Systems like VarifocalReader make use of multi-level summaries to facilitate skimming of large documents~\cite{kochVarifocalReaderInDepthVisual2014}. GistVis incorporates summary visualizations directly in the text allowing readers to search directly for interesting sections~\cite{zou2025gistvis}.

To support this natural skimming behavior, \retrace draws upon a toolkit of principles from HCI and Visualization. 
A classic, overarching principle is that of Shneiderman's mantra: ``Overview first, zoom and filter, then details-on-demand''~\cite{shneiderman2003eyes}. This has been used in mixed-initiative user interfaces~\cite{horvitzPrinciplesMixedinitiativeUser1999}, as well as in progressive disclosure, where information is revealed gradually to manage complexity~\cite{norman1986User}. It is also the guiding principle behind focus+context and overview+detail techniques that show details while maintaining surrounding orientation~\cite{card1999readings, cockburnReviewOverview+detailZooming2009}. These types of strategies are helpful in \textit{sensemaking} for information visualization~\cite{lee2015people}, and are tried-and-true methods for visualization design across many fields.

Recent work has applied these ideas to LLM outputs in various ways. Exploratory interfaces from~\citet {geroSupportingSensemakingLarge2024} show that with support for skimming, a wide variety of sensemaking tasks can now be made tractable. Tools like ChainForge facilitate the evaluation of many prompt variations~\cite{arawjoChainForgeVisualToolkit2024}. To manage the depth of single outputs, systems like Graphologue and Sensescape transform the final text into interactive, multi-level networks~\cite{jiangGraphologueExploringLarge2023a, suhSensecapeEnablingMultilevel2023}.~\citet{liReasonGraphVisualisationReasoning2025} shows preliminary works on how to visualize the reasoning of LLMs by instructing them to generate their answer together with a parse-able structure for visualization. Our work builds directly on this network representation, aiming to transform reasoning traces from a static collection of text into an active scaffold for the cognitive ``sensemaking loop,'' where users forage for cues, build a mental model, and develop insights~\cite{pirolliSensemakingProcessLeverage}.

Our approach also draws from the human-centered explainable AI paradigm, which argues that the value of an explanation lies in a user's ability to understand and act upon it~\cite{amershiGuidelinesHumanAIInteraction2019}. This perspective necessitates designing for a broad audience, acknowledging that AI literacy, the competencies needed to evaluate AI critically, varies widely and influences how explanations are perceived~\cite{longAILiteracyFinding2023, ehsanWhoXAIHow2024}. In line with this, we designed for and evaluated with non-experts to ensure \retrace's effectiveness in supporting sensemaking without requiring specialized AI knowledge.
{We note however, that \retrace makes no observation to the accuracy of the reasoning trace as a model of the underlying, mechanical processes performed by the LLM.}

\retrace{} synthesizes these various ideas into an interactive system tailored to the specific sensemaking challenges posed by LRM reasoning traces. By focusing on the verbose, sequential, and structured nature of the \textit{pre-answer reasoning trace}, our work provides a novel tool for process auditing.

\subsection{Alternative Models and Visualization Systems} 
The sheer verbosity of reasoning traces, which can span thousands of tokens~\cite{marjanovicDeepSeekR1ThoughtologyLets2025}, makes manual inspection for sensemaking or auditing prohibitively time-consuming. These challenges necessitate automated approaches that can parse, structure, and summarize the trace content. Given that the input is unstructured text, recent work has shown the potential of using an LLM-driven pipeline for such visualization tasks. Systems like LIDA can automatically generate visualizations from natural language descriptions of data~\cite{dibiaLIDAToolAutomatic2023}. More specifically, systems like Graphologue convert the final textual outputs of LLMs into interactive semantic graphs to aid comprehension~\cite{jiangGraphologueExploringLarge2023a}, though neither LIDA nor Graphologue are designed for reasoning traces. These systems establish a precedent for using LLMs to bridge the gap between complex text and structured visual representation. 

To visualize the structured output from our pipeline, we draw upon established idioms for sequential and process-oriented data. A natural model for event-based data is the directed acyclic graph (DAG). However, not all visual encodings for DAGs are suited to the specific characteristics of LRM traces. Sankey Diagrams, introduced to depict energy flows~\cite{schmidt2008sankey}, are commonly used to visualize such data, emphasizing edges to identify dominant paths and bottlenecks. LRM traces rarely have diverging paths, prompting us to explore other encodings. Similarly, while storyline visualizations are a popular style for drawing event-based data, they are designed for scenarios in which multiple characters or agents interact over time~\cite{arendt2017matters,tanahashi2012design}, whereas LRM reasoning data involves a single agent.

Our approach finds closer parallels in visualizations of program execution and business processes, though with key distinctions. Control flow graphs (CFGs) model program execution paths, with nodes representing code blocks and edges representing potential execution flow~\cite{ferrante1987program}. While several systems visualize these graphs with a focus on compact layouts and clarity for debugging~\cite{devkota2020ccnav,balmas2004displaying,toprak2014lightweight}, they typically hide the semantic content within a code block, which is the information \retrace emphasizes. Likewise, process mining visualizations~\cite{van2011process,hao2006business} aim to extract insights from real-world business processes but rarely show the rich accompanying text for each step. As \retrace is designed to facilitate understanding of the reasoning text directly, our concerns differ. Provenance in visualization refers to the origin, history, and transformation path of data, which is important for establishing trust and reproducibility~\cite{ragan2015characterizing,xu2020survey}. In this view, \retrace serves as a tool for visualizing the provenance of an LRM's final answer, tracking the chain of reasoning.

The availability of reasoning traces offers a new lens for model interpretability, creating a distinction between high-level, symbolic explanations and low-level, mechanistic ones. Much work has focused on the mechanistic interpretations, with tools providing detailed visualizations of internal model states. For instance,~\citet{vigVisualizingAttentionTransformerBased2019} and~\citet{pierseTransformersInterpret2021} developed tools to visualize attention mechanisms in transformers, revealing token-level interactions. More recently,~\citet{healyFrames_of_mindAnimatingR1s} animated the internal ``thought'' vectors of the DeepSeek-R1 model. Although these low-level approaches provide valuable insights into model mechanics, they do not facilitate the human sensemaking of the high-level, semantic argument constructed in the reasoning trace. Complementary work visualizes aggregate reasoning dynamics as abstract “landscapes,” revealing macro-level model behaviors without exposing step-level content \cite{zhou2025landscape}. Concurrent work has also explored visualizing aggregate reasoning dynamics as abstract ``landscapes'' to reveal macro-level model behaviors \cite{zhou2025landscape}. In contrast, high-level approaches analyze the structure and content of the trace itself. Work by~\citet{marjanovicDeepSeekR1ThoughtologyLets2025} demonstrates that these traces contain a rich, discernible structure and a taxonomy of reasoning patterns that are conducive to explainability.

\begin{figure}
    \centering
    \includegraphics[width=1\linewidth]{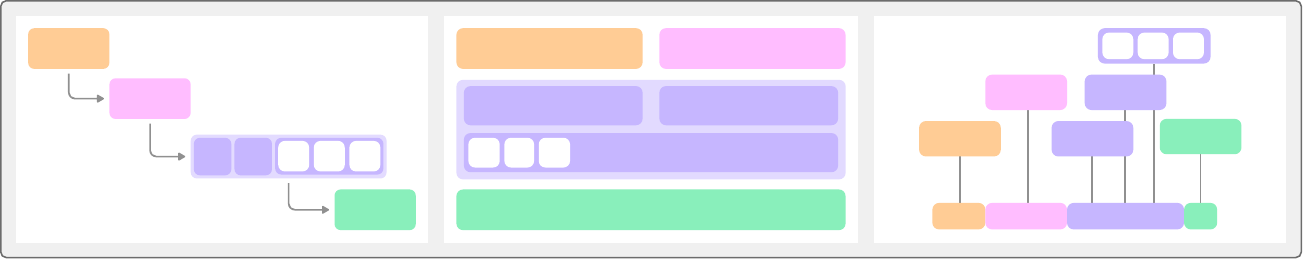}
    \caption{The three early prototypes explored in our exploratory study. From left to right: a Node-Link diagram emphasizing relational flow, Space-Filling Nodes emphasizing hierarchical structure, and a Sequential Timeline emphasizing chronological order.}
    \Description{Three early prototype visualizations of the reasoning trace shown side-by-side. The first, on the left, is a node-link diagram with several nested boxes connected by arrows to show relational flow. The second, in the center, is a Space-Filling diagram composed of nested colored rectangles to show hierarchy. The third, on the right, is a sequential Timeline with colored nested blocks arranged horizontally to show chronological order.}
    \label{fig:prototypes}
\end{figure}

We understand \retrace as an extension of context-based automatic visualization techniques. In this work, we extend previous research in the field of automatic visualization by using LLMs to capture, group, extract and summaries the internal structure of reasoning traces.

\section{Exploratory Study}
We began with three early visualizations to probe the design space. Building on DeepSeek-R1’s inherent phase structure
(Section~\ref{sec:reasoning-trace-structuring}), we manually segmented traces into phases and subphases to serve as ground truth for prototypes. These groups served as the structural building blocks for our prototypes. We instantiated complementary metaphors: a Node–Link diagram (Fig.~\ref{fig:prototypes}, left) to foreground relational flow, Space-Filling Nodes (center) to emphasize hierarchical structure and overview, and a Sequential Timeline (right) to preserve explicit chronology and proportional span. Each prototype supported progressive disclosure and linked summaries back to verbatim steps, which we then evaluated in the exploratory study.

To test our initial designs and guide refinement, we ran an exploratory first-use study with four experts. Three participants self-identified as ML researchers \& engineers, and one as a visualization researcher. We compared the three prototypes against a raw-trace baseline, counterbalancing presentation order (Fig.~\ref{fig:prototypes}). Each expert first explored different reasoning traces in every interface, then completed the same task set later used in the main study. Following, we conducted individual, semi-structured interviews covering layout, color, clarity, summaries, cognitive workload, and desired changes. We transcribed sessions and performed a thematic analysis, clustering quotations into themes. The results revealed convergent insights that shaped the design challenges and goals below.

\subsection{Design Challenges}
Our analysis revealed four insights into key design challenges (C1-C4) :
\begin{description}[font=\bfseries,labelsep=0.6em,leftmargin=!,labelwidth=1.5em]
  \item[C1] \textbf{Maintaining explicit sequence.} 
    Three experts stressed the need for an unambiguous reading order that makes ``\textit{what comes next''} obvious at a glance. One expert preferred the Sequential Timeline's left-to-right scan for quickly judging the process: \textit{``after a glimpse I can know to what percentage it is a thinking process.''} Another highlighted the Space-Filling Nodes' top-to-bottom readability as: \textit{``intuitive because it reads from top to bottom […] it's very clear where to follow the path.''}
  \item[C2] \textbf{Unstructured detail increases cognitive load.} 
    Another central theme, mentioned by three participants, was the dense display of information inside the raw trace and the cognitive demand thereof. One expert noted, \textit{``[In the raw trace] there was everything and you had to scroll […] with other visualizations one can… just show the stuff that you wanted to see.''} 
  \item[C3] \textbf{Experts work overview-first and drill down selectively.}
    All four participants consistently adopted a top-down search strategy inside our prototypes, enabled by summaries and multi-level detail. As one expert described their process: \textit{``I will first read the summarization and the label  [then] narrow my search space.''} Another stated, \textit{``I usually will look for the summaries first… then I would just not click the other small boxes.''} 
  \item[C4] \textbf{Surface reasoning strategy variants and re-thinking.}
    Furthermore, three of the experts noted, the utility of being able to distinguish and compare different reasoning phases within the trace. For example, \textit{``I can just directly compare [the outputs of different approaches].''} Another desired the ability to \textit{``Differentiate between replication and trying alternatives.”}
\end{description}

\subsection{Design Goals}
We mapped the challenges to three design goals that guided the development of ReTrace:
\begin{description}[font=\bfseries,labelsep=0.6em,leftmargin=!,labelwidth=1.5em]
  \item[G1] \textbf{Make sequence explicit.} 
    Use a layout that preserves the sequential nature of the trace, with clear step numbering and unambiguous ordering. (C1)
  \item[G2] \textbf{Offload cognitive work with structure.}
    Provide concise step summaries and meaningful phase labels that chunk the reasoning trace, surface the primary strategy, and reduce cognitive load. (C2, C4)
  \item[G3] \textbf{Multi-level detail on demand.}
    Support rapid exploration through progressive disclosure—from phase overviews to subphase summaries to raw step content and evidence—so readers can skim, zoom, and drill down as needed. (C3)
\end{description}

\subsection{Pilot Adjustments}
Based on exploratory feedback, we made several adjustments. We added a legend for color semantics, kept expanded items in the foreground, limited simultaneous expansions, and moved the distribution view to a separate display. We also removed the Node–Link diagram from the main study. Although one expert valued its arrow guidance, other participants preferred the explicit chronology of the Sequential Timeline and the structured overview of Space-Filling Nodes. In dense traces, link crossings and spatial search offered no advantage over clear left-to-right or top-to-bottom reading, so we advanced the other two prototypes.

\section{ReTrace Design \& Implementation}
Informed by the design goals derived from our exploratory study, we present \retrace, an interactive visualization for LRM reasoning traces. Our system structures the “think out loud” LRM reasoning with an LLM-supported pipeline and then visualizes it. Below we describe the design and implementation of \retrace and link the respective design goals (D1-D3).

\begin{table}[htbp]
\centering
\small
    \caption{The reasoning taxonomy used in \retrace. The table shows the four primary reasoning phases (e.g., Problem Definition \& Scoping), their inherent subphases (e.g., Rephrase), along with definitions and illustrative examples from reasoning traces.
    }
    \Description{A two-column table defining the reasoning taxonomy used in the system. The table is organized into four primary phases: ``Problem Definition and Scoping'', ``Initial Solution and Exploration'', ``Iterative Refinement and Verification'', and ``Final Decision''. Each phase contains several subphases, such as 'Rephrase' or 'Correction'. For each subphase, the table provides a definition and illustrative textual examples from reasoning traces, like ``Let me correct that...''.}
    \label{tab:taxonomy}
\begin{tabularx}{\linewidth}{@{}l l >{\raggedright\arraybackslash}X@{}}
\toprule
\multicolumn{3}{l}{\textbf{Problem Definition \& Scoping}} \\
\cmidrule(lr){1-3}
& Rephrase & \textit{Restates the task to confirm understanding — examples: ``Let me restate the problem\ldots''; ``Basically, the goal is\ldots''} \\
\cmidrule(lr){2-3}
& Define Goal & \textit{Specifies the required objective/output — examples: ``I need to find\ldots''; ``The aim is to\ldots''} \\
\cmidrule(lr){1-3}
\multicolumn{3}{l}{\textbf{Initial Solution \& Exploration}} \\
\cmidrule(lr){1-3}
& Decomposition \& Execution & \textit{Breaks the problem into steps and proceeds — examples: ``First, I should\ldots''; ``Let's break this down\ldots''} \\
\cmidrule(lr){2-3}
& First Answer & \textit{States a complete initial result — examples: ``So, the answer is\ldots''; ``Therefore, the result is\ldots''} \\
\cmidrule(lr){2-3}
& Confidence Qualification & \textit{Gives a quick plausibility check — examples: ``Hm, let me verify that\ldots''; ``Does that make sense?\ldots''} \\
\cmidrule(lr){1-3}
\multicolumn{3}{l}{\textbf{Iterative Refinement \& Verification}} \\
\cmidrule(lr){1-3}
& Pausing to Rethink & \textit{Stops to reconsider direction — examples: ``Wait\ldots''; ``Hold on\ldots''} \\
\cmidrule(lr){2-3}
& Correction & \textit{Fixes assumptions/calculations — examples: ``Let me correct that\ldots''; ``Let me recalculate that part\ldots''} \\
\cmidrule(lr){2-3}
& Re-examine & \textit{Re-checks prior work without progress — examples: ``Let me check that again\ldots''; ``Let me review the steps\ldots''} \\
\cmidrule(lr){2-3}
& Try Alternative & \textit{Explores a different approach — examples: ``Alternatively\ldots''; ``Another way to see this is\ldots''} \\
\cmidrule(lr){2-3}
& Abandonment & \textit{Drops an unproductive line — examples: ``This approach is a dead end\ldots''; ``I'll drop this path and try another\ldots''} \\
\cmidrule(lr){1-3}
\multicolumn{3}{l}{\textbf{Final Decision}} \\
\cmidrule(lr){1-3}
& Stating Confidence & \textit{Expresses certainty in the answer — examples: ``I'm confident this is correct\ldots''; ``I think this is right now\ldots''} \\
\cmidrule(lr){2-3}
& Preparing Output & \textit{Formats and delivers the result — examples: ``So, the final answer is\ldots''; ``Here's the result:\ldots''} \\
\bottomrule
\end{tabularx}

\end{table}

\subsection{Reasoning Trace Structuring}\label{sec:reasoning-trace-structuring}
\begin{figure}
    \centering
    \includegraphics[width=1\linewidth]{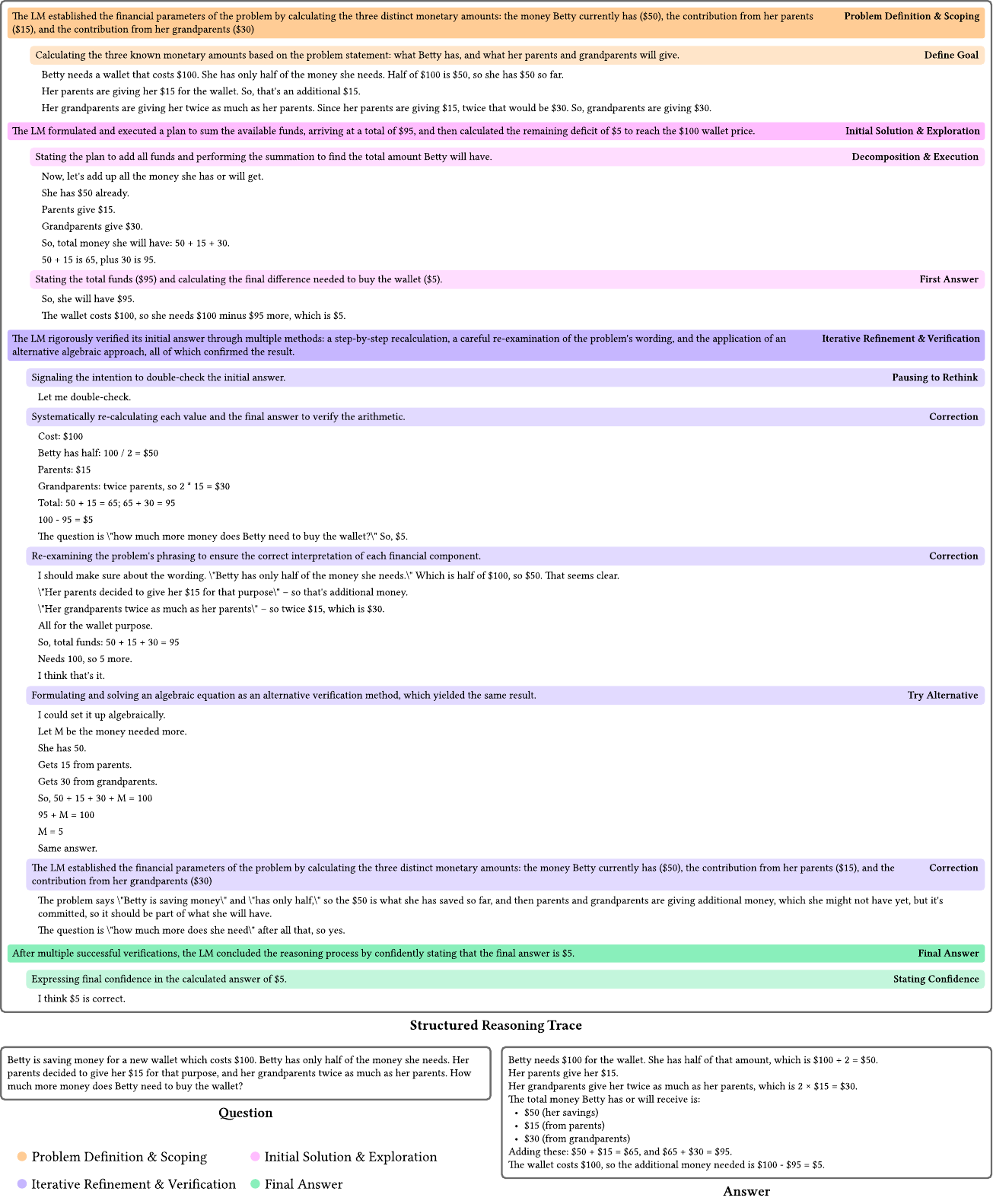}
    \caption{
    Example of a structured reasoning trace on a GSM8K sample, annotated by \retrace. Colored annotations show LLM-generated summaries, indentation groups, and labels' reasoning taxonomy. Question from GSM8K bottom left, LRM generated answer bottom right. Legend shows assignment to the respective reasoning phase.
    }
    \Description{An example of a reasoning trace after it has been structured and annotated by ReTrace. The text of the reasoning trace is grouped by indentation and colored backgrounds corresponding to different reasoning phases. Each group is headed by an LLM-generated summary. A legend at the bottom maps the four background colors to their respective reasoning phases: Problem Definition and Scoping, Initial Solution and Exploration, Iterative Refinement and Verification, and Final Answer. Two boxes at the bottom show the original input question to the LRM from the GSM8K dataset and the LRM generated answer to this question.}
    \label{fig:trace-example}
\end{figure}

To structure raw reasoning traces, we utilize a validated four-phase taxonomy for DeepSeek-R1~\cite{marjanovicDeepSeekR1ThoughtologyLets2025}. 
This taxonomy distinguishes between four consecutive phases, relabeled in this work to make them accessible and understandable without prior knowledge. 
The reasoning always starts with a Problem Definition, rephrasing the problem and defining the goal, followed by an Initial Solution \& Exploration phase, decomposing the task and generating an initial solution. An optional Iterative Refinement \& Verification phase uses various ways of corrections, re-examinations, or alternatives to validate the reasoning, which concludes in a Final Answer phase, in which the model prepares its confidence and response. 
Marjanović et al.~\cite{marjanovicDeepSeekR1ThoughtologyLets2025}
validated the source taxonomy with human annotation and cross-checked against LLM-assisted structuring, establishing reliability for this phase vocabulary. We expose recurring subphases via a second level to make them explicit (G3). 
Each phase can be recognized not only by its content but also by recurring phrases (Tab.~\ref{tab:taxonomy}). Informed by this taxonomy, we instruct an LLM to generate a structured trace from the raw input. Instructions to the LLM also define that each phase can only be a group of consecutive steps (G1). In addition to the grouping, we also generate summaries for each main and subphase (G2) (Fig.~\ref{fig:trace-example}).

\subsection{Reasoning Trace Visualization}
The interaction design for both of our proposed designs are informed by G3 (Multi-level detail on demand). At a high level, each design supports: 

\begin{itemize}
    \item[] \textbf{Canvas Navigation:} Visualizations are rendered on a large canvas supporting standard zoom and pan.
    \item[]  \textbf{Progressive Disclosure:} Drill-down interactions allow one to go from a high-level overview to granular step-by-step details.
    \item[]  \textbf{Summary view:} Each design keeps at all times in view a visual summary of the relative number of phases, either in a separate bar chart for the Space-Filling Nodes or integrated directly into the visualization for the Sequential Timeline.
\end{itemize}
Two complementary aspects of reasoning are their hierarchical organization and their sequential progression. To capture these, we design two visualizations: the Space-Filling Nodes highlight hierarchical structure (G2) while the Sequential Timeline highlights the chain of thought (G1). 

\begin{figure}[t]
    \centering
    \includegraphics[width=1\linewidth]{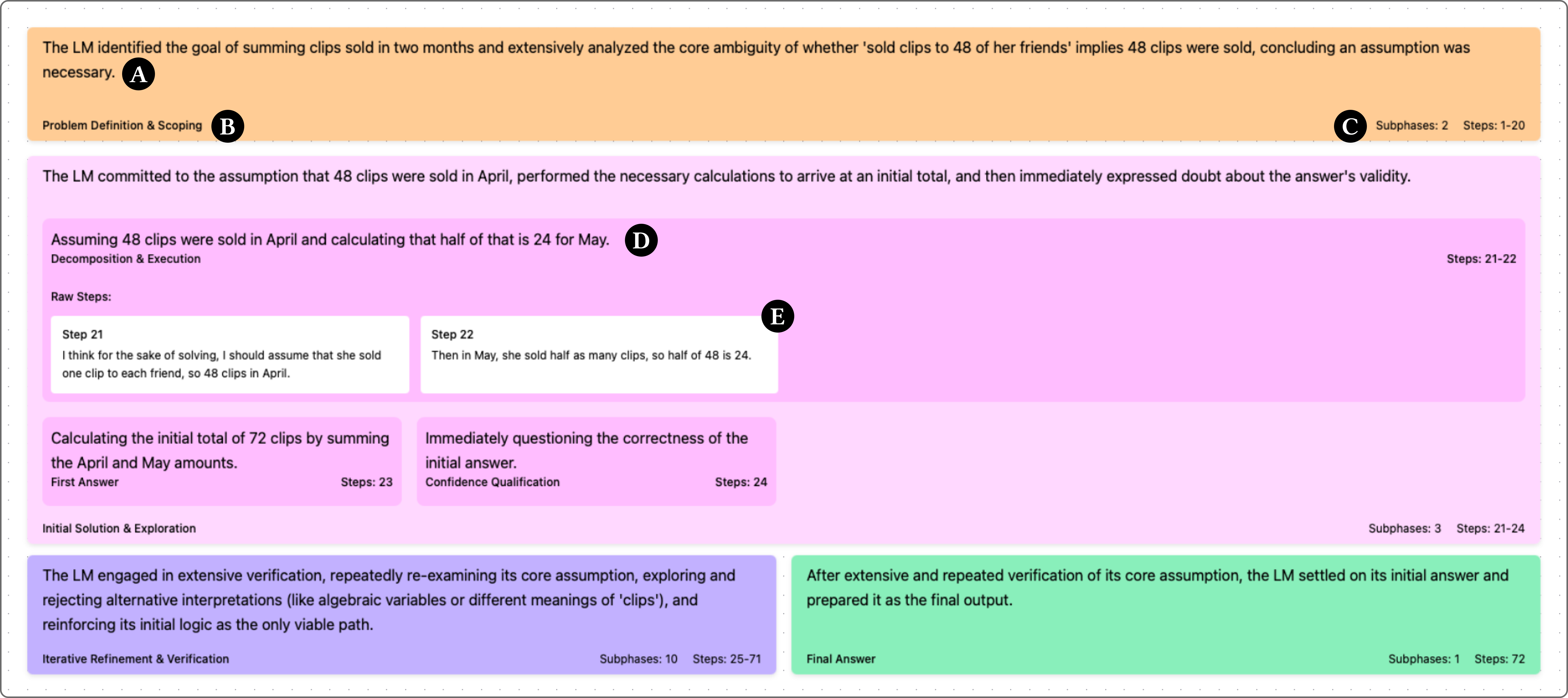}
    \caption{The Space-Filling Nodes visualization in the \retrace interface. This view organizes the reasoning trace hierarchically. The main panels represent primary reasoning phases, each showing (A) a summary, (B) its category label, and (C) the count of subphases and steps. Clicking a main panel reveals (D) nodes for each subphase, which can be further expanded to show (E) the raw text of individual reasoning steps.  
    }
    \Description{Screenshot of the Space-Filling Nodes hierarchical visualization. The interface is composed of large, colored rectangular panels representing the primary reasoning phases. Each panel displays a summary, a category label, and step counts. The screenshot demonstrates the drill-down interaction: one main panel is shown expanded to reveal smaller nodes for its subphases, and one of those subphase nodes is further expanded to show the original raw text of individual reasoning steps.}
    \label{fig:Space-Filling-ui}
\end{figure}

\subsubsection*{Space-Filling Nodes} \label{sec:design-space}
Given that DeepSeek-R1's reasoning traces almost always follow a four part structure, we first consider a design that at all times fills the screen space; we call this the Space-Filling Nodes design. This design also utilizes the hierarchical structure of the data to create a focus+context navigation. This treemap-style visualization gives a quick at-a-glance overview while enabling progressive drill-down for finer granularity. Initially, the highest level of the taxonomy hierarchy is shown as a colored block (one in each quadrant). Each block represents a navigable summary, with a short one sentence description of the reasoning that took place in that phase; see Fig.~\ref{fig:Space-Filling-ui} (A). To support user orientation, additional details are shown in the block: a category label, the total number of subphases, and the range of raw reasoning steps contained within the block, all shown in Fig.~\ref{fig:Space-Filling-ui} (B,C).

When a block is clicked, it expands to reveal its internal structure, and other blocks are resized and moved to accommodate the selected block's expansion. Inside the selected block are representations of the second level of the hierarchy (subphases) as smaller blocks. Each of these internal blocks model its parent, with a short summary and additional information. Finally, drilling further down in a subphase block reveals the raw text of the original reasoning trace. 

Despite the strong emphasis on the hierarchy, there are several potentially desirable attributes the Space-Filling Nodes do not capture. Firstly, all categories in the reasoning trace are depicted as the same size, regardless of whether or not they are of equal proportion. This may induce bias in the perceived importance of each step. There is additionally a loss of perceived sequential flow. Although the step ranges are annotated and the trace always has the same structural pattern, the decoupling of the sequence from the visual encoding makes it more difficult to reconstruct the precise order of thought.

\begin{figure}[t]
    \centering
    \includegraphics[width=1\linewidth]{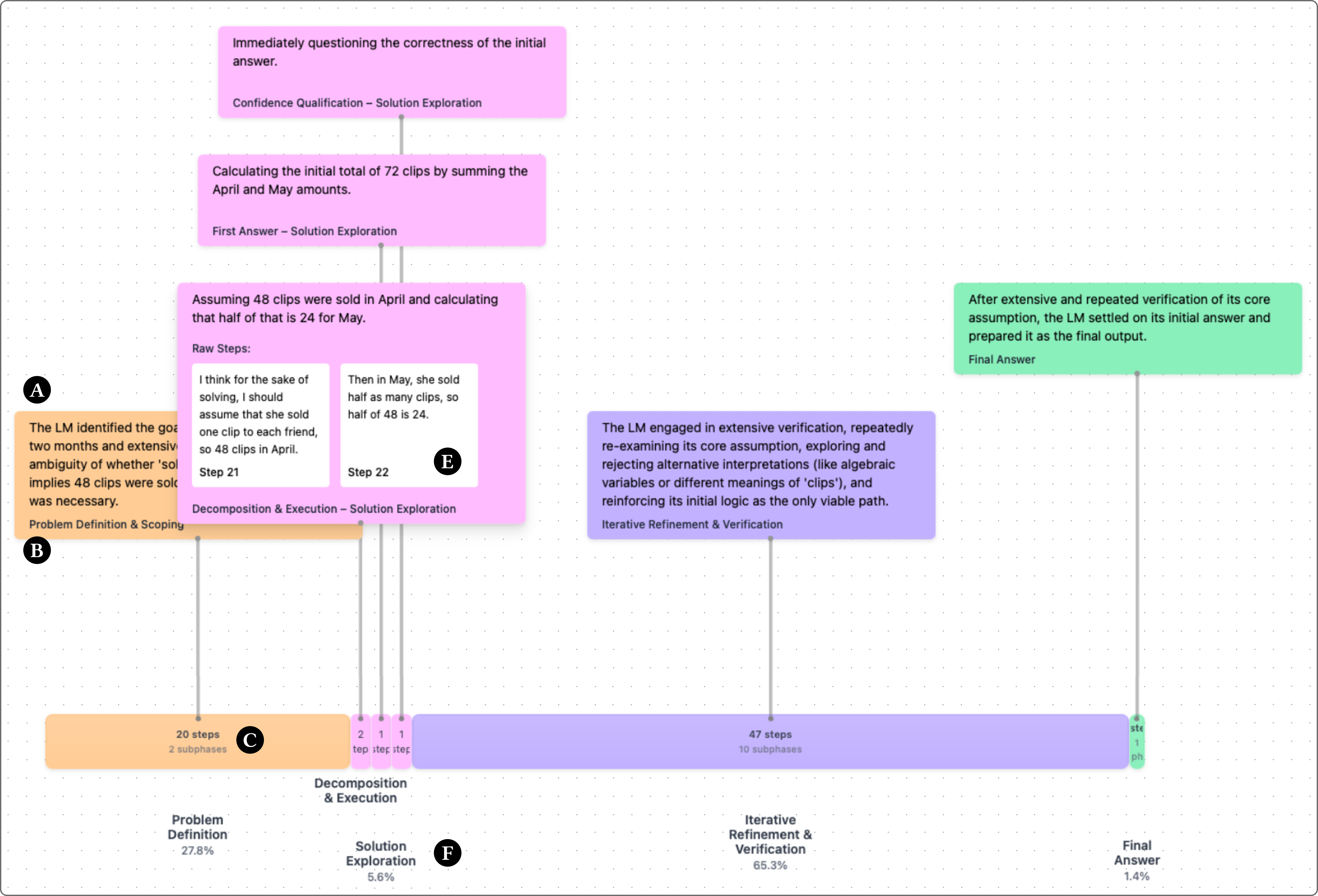}
    \caption{The Sequential Timeline visualization in the \retrace interface. This view presents the reasoning trace chronologically. Each phase is represented by a box containing (A) a clickable summary and (B) its category label. The timeline bar below (C) shows the count of subphases and steps in proportional length of each reasoning phase. Expanding a phase reveals (D) nodes for subphases, which can be expanded to show (E) the raw text of individual steps. The category label at the bottom (F) shows the share of steps for the main phases.
    }
    \Description{Screenshot of the Sequential Timeline chronological visualization. At the bottom of the interface, a horizontal bar is segmented and colored to show the proportional length of each reasoning phase. Above this bar, boxes for each phase contain a summary and can be clicked to expand. The screenshot shows one phase expanded to reveal nodes for its subphases, one of which is further expanded to show the original raw text of the reasoning steps, demonstrating the drill-down interaction.}
    \label{fig:timeline-ui}
\end{figure}

\subsubsection*{Sequential Timeline} \label{sec:design-time}
In contrast to the Space-Filling Nodes design, which emphasizes the hierarchical structure, the Sequential Timeline view emphasizes instead the temporal progression of the reasoning trace. Phases are laid out along the horizontal axis, with their length encoding the number of underlying steps within them. This design is particularly well suited for understanding the proportion of time spent across phases and for mentally reconstructing the exact chain of thought (Fig.~\ref{fig:timeline-ui}). 

Each phase initially appears as a color coded section of a bar across the bottom of the screen. In the middle of each bar a link traces up to a box containing a one sentence summary of the reasoning that occurred in that phase. Clicking on this box reveals the subphases contained within the phase, and subdivides the horizontal timeline bar accordingly. The original text can be found by clicking within a subphase.
This representation addresses several of the potential limitations of the Space-Filling Nodes. Proportional visual scaling of the phases on the timeline bar mitigates the bias in their perceived importance. Likewise, sequential order is explicit, reducing the cognitive effort needed to reconstruct the order of thought. 

On the other hand, the Sequential Timeline has its own limitations. The emphasis on the temporal progression comes at the cost of obscuring the global hierarchy of the phases and subphases, so a user may lose a sense of the higher order taxonomy. A second shortcoming is that as phases often vary dramatically in length, long spans may dominate the visual space relegating short but important phases to a small section of the screen which may be easy to overlook or difficult to click. 

Taken together, the Space-Filling Nodes and Sequential Timeline views illustrate to complementary visualization styles in the design space of reasoning trace visualization. While the Space-Filling Nodes emphasize the hierarchical structure and navigability, the Sequential Timeline emphasizes the temporal sequence and balance between phases. Interestingly, the potential limitations of each style is addressed with the other. It is not clear which style users may prefer or which style may be more effective. For this reason, we incorporate both designs into our study, to compare against a raw text baseline, and against each other.

\begin{figure}[t]
    \centering
    \includegraphics[width=1\linewidth]{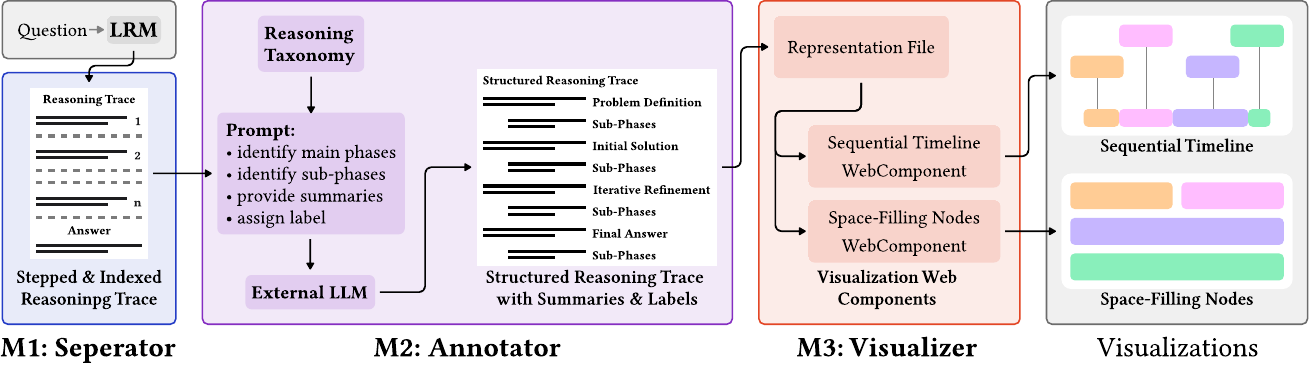}
    \caption{The architecture of the \retrace data processing pipeline. The pipeline consists of three modules: (M1) Separator, which indexes the raw reasoning trace; (M2) Annotator, which uses an external LLM and a reasoning taxonomy to structure, summarize, and label the trace; and (M3) Visualizer, which generates the interactive web components for the two visualization types.
    }
    \Description{Architecture diagram of the ReTrace data processing pipeline, showing a flow through three main modules labeled M1, M2, and M3. Module 1, Separator, takes the raw reasoning trace, from an LRM output, separates and indexes it into steps. Module 2, Annotator, uses an external Large Language Model and a reasoning taxonomy to structure and summarize the stepped trace. Module 3, Visualizer, uses this structured data to generate the two interactive visualizations with web components. Those two visualizations are the Sequential Timeline and the Space-Filling Nodes.}
    \label{fig:retrace-pipeline}
\end{figure}

\subsection{Implementation}
\retrace was implemented and deployed as a web application. We used the Svelte framework\footnote{https://svelte.dev} with the Svelte Flow library\footnote{https://svelteflow.dev} to build a frontend and Python for our backend. To generate the reasoning traces, we prompted \texttt{DeepSeek-R1}~\cite{deepseek-aiDeepSeekR1IncentivizingReasoning2025a} via its official API, as it is one of the few available LRMs that reveals its complete reasoning trace to the users. For the structuring and annotation of reasoning traces, we utilized \texttt{gemini-2.5-pro}~\cite{comaniciGemini25Pushing2025}. Our end-to-end pipeline consists of three modules (Fig.~\ref{fig:retrace-pipeline}), described below.

\begin{itemize}
    \item[] \textbf{Separator Module (M1)}: Receiving the complete output from the LRM, \retrace begins with dividing the reasoning trace into indexed steps. We extract the reasoning trace from the API response and separate it with the newline character \texttt{\textbackslash n} as a delimiter. 
    
    \item[] \textbf{Annotator Module (M2)}: With the stepped trace as input, we use our taxonomy (Tab. ~\ref{tab:taxonomy}) to annotate the reasoning trace. We instruct the external LLM, according to Section~\ref{sec:reasoning-trace-structuring}, and check the response for correctness of our JSON schema. In a sub-step, we combine the response from the LLM containing the groups with indexes, labels, and summaries, and parse the verbatim steps from the stepped trace back to complete our structured reasoning trace.
    
    \item[] \textbf{Visualizer Module (M3)}: Finally, we parse our structured reasoning trace, annotated with all steps, into a visualization-agnostic representation file. Both of \retrace's visualizations web components can access this file to display them in our user interface.

\end{itemize}

\section{User Study}
To evaluate the utility and effectiveness of \retrace in making LRM reasoning traces more understandable, we conducted a within-subjects usability study. We compared our two interactive visualizations, Space-Filling Nodes and Sequential Timeline, against a text-based baseline, measuring their impact on participant comprehension, trust, and perceived cognitive workload.

The study, described below, was designed to explore the following research questions:
\begin{description}[font=\bfseries,labelsep=0.6em,leftmargin=!,labelwidth=2.0em]
  \item[RQ1] Does \retrace improve comprehension of model reasoning relative to a Raw Trace, and how does this differ between the Sequential Timeline and Space-Filling Nodes?
  \item[RQ2] Does \retrace reduce workload and task time and increase perceived ease, and how does this differ between the Sequential Timeline and Space-Filling Nodes?
  \item[RQ3] To what extent does \retrace support calibrated trust and confidence in model answers and perceived logical soundness, and how does this differ between the Sequential Timeline and Space-Filling Nodes?
\end{description}

\subsection{Study Design}
At a high level, the study proceeded by showing participants one of three different visualization styles of an LRM reasoning trace and asking them various questions. 
We implemented the study interface using ReVISit~\cite{dingReVISitSupportingScalable2023}, a web-based framework for evaluating interactive visualizations. This allowed us to embed the live \retrace system directly within the study instrument, presenting both the visualization and the task questions in a single browser view (Fig.~\ref{fig:study-ui} 
), automated assignment of participants to counterbalanced experimental conditions, and robust logging.
Below we describe the conditions, materials and questions of our study.
.

\subsubsection*{Conditions}
Our study compares the visualizations provided by \retrace to the raw text baseline. We follow a within-subjects study design with three experimental conditions, varying the presentation of the reasoning trace in the main interface view. We randomized and counterbalanced the order of conditions and datasets using a balanced Latin square to limit order effects. In every condition, a collapsible bottom bar (``drawer'') was always visible, showing the exact input given to the LRM and its final generated output (Fig.~\ref{fig:study-ui}). All other interface elements, such as typography and color schemes, were held constant to isolate the effects of the visualization.
\begin{itemize}
\item \textbf{Raw Trace (Baseline).} The trace was rendered as plain text in a vertically scrolling panel. Hovering over a line revealed its step index.
\item \textbf{Space-Filling Nodes.} The trace was visualized using a hierarchical, two-column layout (Section~\ref{sec:design-space}). Interaction was via drill-down clicks. A persistent bar chart in the bottom drawer showed the phase distribution.
\item \textbf{Sequential Timeline.} The trace was presented as a left-to-right timeline (Section~\ref{sec:design-time}). Interaction was identical to the Space-Filling Nodes condition. The phase distribution was integrated directly into the timeline's axis.
\end{itemize}

\begin{table}[h]
\caption{
    The three questions used to generate the reasoning trace in our user study. We sourced each problem from a different benchmark dataset (DROP, GSM8K, Zebra) to represent a distinct reasoning challenge: reading comprehension, mathematical problem-solving, and logic puzzles.
    }
    \Description{Table listing the three questions used in the user study. The table shows the full text for three distinct problems sourced from different benchmarks. The first is a reading comprehension task from the DROP dataset about a historical war. The second is a math word problem from the GSM8K dataset about selling clips. The third is a logic puzzle from the Zebra dataset about people in houses.}
    \label{tab:materials}
\centering
\small
\begin{tabularx}{\textwidth}{@{} l >{\raggedright\arraybackslash}X @{}}
\toprule
\textbf{Dataset} & \textbf{Sample} \\
\midrule
DROP  & \textit{The Polish-Lithuanian-Teutonic War or Great War occurred between 1409 and 1411, pitting the allied Kingdom of Poland and Grand Duchy of Lithuania against the Teutonic Knights. Inspired by the local Samogitian uprising, the war began by Teutonic invasion of Poland in August 1409. As neither side was ready for a full-scale war, Wenceslaus IV of Bohemia brokered a nine-month truce. After the truce expired in June 1410, the military-religious monks were decisively defeated in the Battle of Grunwald , one of the largest battles in medieval Europe. Most of the Teutonic leadership was killed or taken prisoner. While defeated, the Teutonic Knights withstood the siege on their capital in Marienburg and suffered only minimal territorial losses in the Peace of Thorn . Territorial disputes lasted until the Peace of Melno of 1422. However, the Knights never recovered their former power and the financial burden of war reparations caused internal conflicts and economic decline in their lands. The war shifted the balance of power in Eastern Europe and marked the rise of the Polish-Lithuanian union as the dominant power in the region. — How many years between the truce expiring and the Peace of Melno?} \\
\midrule
GSM8K & \textit{Natalia sold clips to 48 of her friends in April, and then she sold half as many clips in May. How many clips did Natalia sell altogether in April and May?} \\
\midrule
Zebra & \textit{There are 4 houses, numbered 1 to 4 from left to right, as seen from across the street. Each house is occupied by a different person. Each house has a unique attribute for each of the following characteristics:\newline - Each person has a unique name: Arnold, Alice, Peter, Eric\newline - People have unique favorite book genres: mystery, romance, science fiction, fantasy\newline \#\# Clues:\newline 1.Arnold is the person who loves science fiction books. 2.The person who loves fantasy books is not in the second house. 3.Alice is not in the second house. 4.The person who loves romance books is in the first house. 5.Peter is the person who loves romance books. 6.Arnold is somewhere to the right of Alice.\newline What is BookGenre of the person who lives in House 2?\newline What is Book Genre of the person who lives in House 2?} \\
\bottomrule
\end{tabularx}

\end{table}

\subsubsection*{Materials}:
The stimuli for the study consisted of three reasoning traces generated from different benchmark datasets. 
GSM8K (math word problems)~\cite{cobbeTrainingVerifiersSolve2021}, DROP (reading comprehension)~\cite{duaDROPReadingComprehension2019}, and a ZebraLogic (logic puzzles)~\cite{linZebraLogicScalingLimits2025}. We selected one problem from each dataset (Tab.~\ref{tab:materials}). We chose these to represent a diverse set of challenging tasks with verifiable ground-truth answers. 

From each dataset, we prompted our LRM to produce a complete reasoning trace. \retrace the processes the raw traces (Section~\ref{fig:retrace-pipeline}) to generate the structured data for the visualization conditions. The underlying reasoning content was identical across all conditions for a given problem. The traces varied in structure: the \textit{GSM8K} and \textit{DROP} traces were dominated by verification/refinement (65.3\% and 88.3\% of steps, respectively), whereas the \textit{Zebra} puzzle showed a more balanced split between exploration (46.3\%) and verification (45.5\%).

\subsubsection*{Study Questions} \label{sec:study-tasks}
\begin{figure}[t]
    \centering
    \includegraphics[width=1\linewidth]{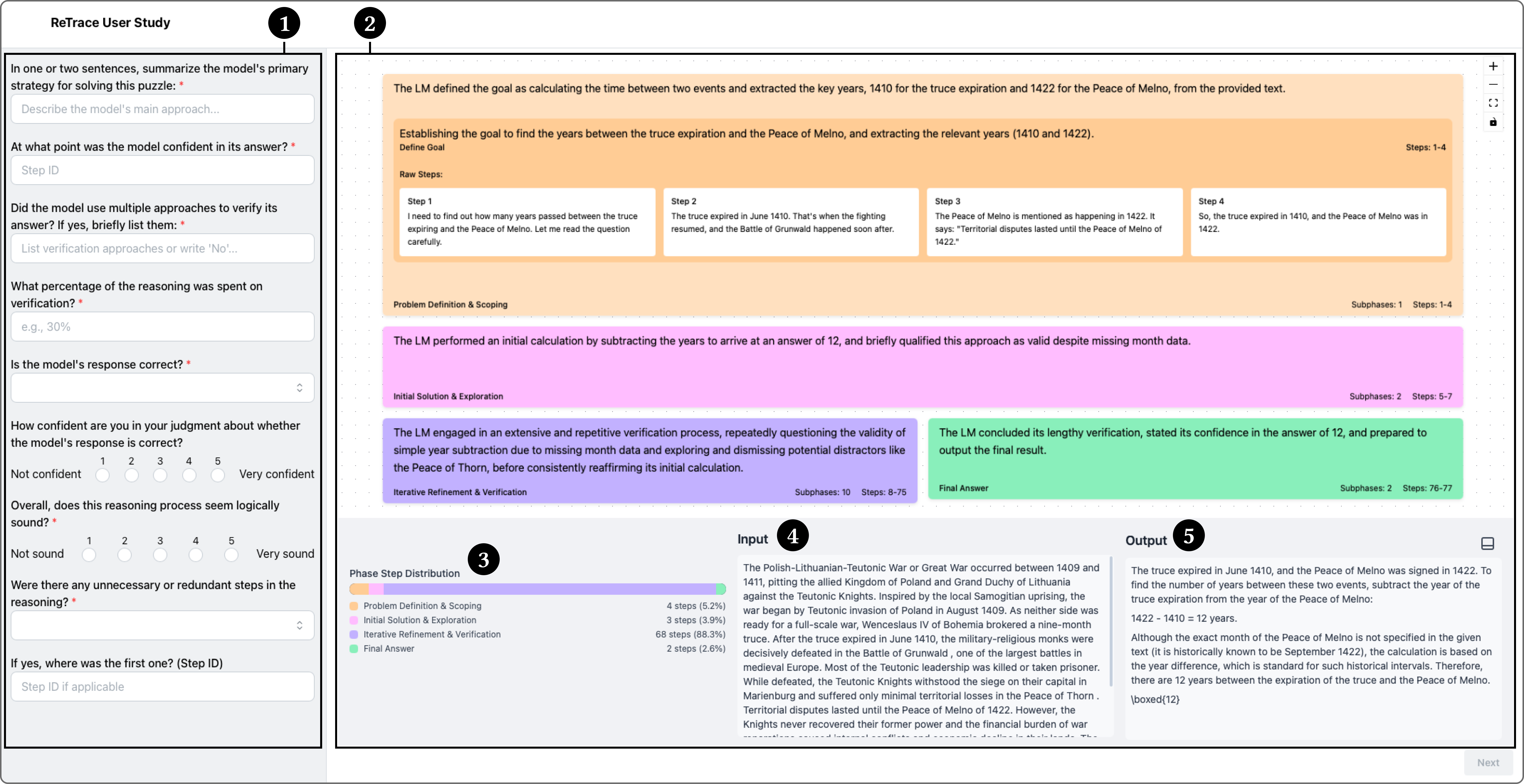}
    \caption{An example of the study interface. The interface consists of (1) a fixed panel on the left displaying task questions and (2) the main view containing one of the three conditions. A collapsible bottom drawer displays (3) a phase distribution bar (only shown for Space-Filling Nodes), (4) the original input question, and (5) the final LRM-generated answer.
    }
    \Description{Screenshot of the user study evaluation interface. The screen is divided into three main parts. On the left is a fixed panel listing the nine task questions for the participant. The main view in the center displays one of the interactive visualization conditions. At the bottom, a collapsible drawer contains the original problem text and the Large Reasoning Model's final generated answer. The condition shown is the Space-Filling Nodes layout, which is the only condition that additionally shows a phase distribution bar.}
    \label{fig:study-ui}
\end{figure}

In each trial, we asked participants to review a single reasoning trace and answer nine evaluation questions derived from our design goals, all shown in the sidebar of Fig.~\ref{fig:study-ui}. To get insights into participant comprehension (RQ1) we asked them to do the following: to summarize the model's primary strategy, to identify the step at which the model became confident in its answer, to decide whether the LRM used verification approaches and, if so, which ones, to estimate the percentage of reasoning devoted to verification, and finally to indicate whether they perceive any steps as unnecessary or redundant, providing the step ID. We designed this set to measure comprehension at multiple depths. The high-level task of summarizing the strategy assesses sensemaking and schema formation~\cite{russellCostStructureSensemaking1993}. In contrast, identifying specific steps and flaws probes the deeper analytical understanding required by user needs in XAI~\cite{liaoQuestioningAIInforming2020}. To assess trust and confidence (RQ3) in the model's reasoning and final answer, we asked participants to evaluate the correctness of the final response. They also rated their confidence in this judgment and provided an overall logical soundness rating. Both measures used a 5-point Likert scale as a measurement for the evaluation of XAI systems~\cite{mohseniMultidisciplinarySurveyFramework2021}. The study interface automatically logged the time spent on task for each trial (Fig.~\ref{fig:study-ui}). 

Following completion of the three trials, participants were asked to respond to questions about the perceived cognitive workload of each condition using the post-task SEQ~\cite{sauro2009seq} NASA-TLX~\cite{hart1988development}. Finally, a semi-structured interview with the participants was conducted  to collect further qualitative feedback.

\subsection{Study Procedure}
All study sessions took place one-on-one over a video conferencing platform and lasted approximately 60 minutes. With participants' consent, we recorded their screens and audio for subsequent analysis. Our institution's internal review board approved this study. To ensure that all participants had the same environment, we asked them to work in a quiet well-lit room, recommended to work on a monitor with at least 27", and to interact with the study using a mouse and keyboard.

\paragraph{Introduction and Exploration} 
Each session began with an informed consent process and a brief introduction. We provided participants with a brief tutorial to familiarize them with the study's objectives and interface. We introduced the concept of LRM reasoning traces and explained how to interact with each of the three conditions using a sample reasoning trace not included in the main experiment. This phase ensured participants understood the functionalities of the baseline view and both \retrace visualizations before the experimental trials began.

\paragraph{Experimental Trials} In our study, participants completed three trials. In each trial, they were presented with one of the three reasoning traces under one of the three interface conditions. Participants were instructed to use the provided interface to answer the nine task questions (as detailed in Section~\ref{sec:study-tasks}). We did not impose a time limit, and participants decided when they were ready to proceed after completing the questions for each trial. Following the completion of the tasks in each trial, participants immediately filled out the post-task SEQ~\cite{sauro2009seq} and NASA-TLX~\cite{hart1988development} questionnaires. 

\paragraph{Post-study questionnaire} After completing the experimental trials, we asked participants to fill out a post-task survey consisting of Likert-scale questions to rate the overall usefulness of the visualizations versus the baseline, state their preferred representation, and provide open-ended justifications for their ratings.

\paragraph{Semi-structured Interview} Following the survey, we interviewed participants with a series of open-ended questions to collect more qualitative feedback about their experience exploring reasoning with our tool. We began by probing how the raw trace vs structured visualization changed their comprehension and confidence in understanding the model's reasoning and the role of interaction in aiding them to answer the tasks. We also asked participants for suggestions for future improvement.

\subsection{Participants}
We recruited 18 participants (7 identified as female, 10 as male; 1 preferred not to identify) via a university-wide mailing list. All participants who registered through our survey subsequently met our criteria for participation. Study sessions took on average 60 minutes, and we compensated each participant with \texteuro15. They came from a diverse range of academic disciplines, including both technical and non-technical fields (Computer Science \& Engineering=11, Humanities=3, Social Science=3, Natural Science=1). On a 5-point Likert scale (1=Novice, 5=Expert), participants reported a medium familiarity with visualizations ($M{=}2.72, SD{=}1.36$) an LLM \& AI reasoning  familiarity ($M{=}2.94, SD{=}1.26$). They also reported using another 5-point Likert scale (1=Never, 5=Multiple times a day), a self-reported medium to high LLM usage frequency ($M{=}3.72, SD{=}1.27$).

\section{Results}
After collecting all responses from all participants, we explore our research questions through quantitative and qualitative analysis of the results. 
For quantitative analysis, we take participant responses for the Likert data and for questions which there is a ground truth answer (Is the model's response correct?, At what point was the model confident?, and What percentage of the reasoning was spent of verification?). For the summary question, we assessed the quality by having two researchers, blind to the condition, code participant responses using a 3-level rubric (0=insufficient, 1=partial, 2=complete) and resolved disagreements through discussion. For the verification questions, we compared the responses to a ground-truth codebook, and again resolved disagreements through discussion. For all pairwise comparisons we used the non-parametric Wilcoxon signed-rank test. To account for multiple comparisons, we applied a Holm-Bonferroni correction with a significance threshold of $\alpha{=}0.05$.

We also collected qualitative data from open-ended survey responses and semi-structured interviews conducted at the end of each session. We then performed a thematic analysis on the interview transcripts from 17 of our 18 participants, as one recording was lost due to a technical problem. Our process involved transcribing the interviews, segmenting the content into quotations via a spreadsheet, and using an inductive analysis to organize these quotations into meaningful themes. This qualitative analysis provides the explanatory context for our findings. We present the results of our study in the following section.

Below, we report and analyze the difference in participants’ study responses when using the two \retrace visualizations and against the baseline. Our interview analysis resulted in six themes (T1-T6), and we present a summary of these in Tab.~\ref{tab:interview-findings}. We referred to participants as P1–P18 in the Results and Discussion.

\subsection{Does ReTrace improve comprehension? (RQ1)}
\begin{figure}[t]
    \centering
    \includegraphics[width=1\linewidth]{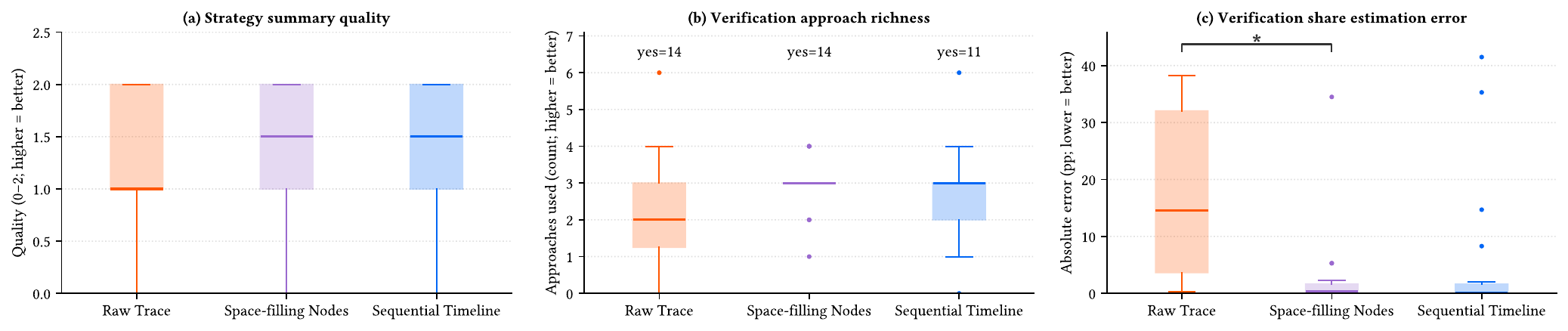}
    \caption{Results for user comprehension (RQ1) across the three task questions by layout. Panels show: (a) quality of participants' strategy summaries (0-2 scale, higher is better); (b) the number of participants identifying verification approaches and respective richness (count; higher is better); and (c) the absolute error in estimating the verification share (lower is better). Boxplots show median and interquartile range (IQR); a star (*) indicates a statistically significant difference ($p_{\text{Holm}}<0.05$).
    }
    \Description{Three panels, each with three boxplots, comparing user comprehension across the Raw Trace, Space-Filling Nodes, and Sequential Timeline conditions. Panel (a), ``Strategy summary quality,'' shows higher median scores for both visualizations than for the Raw Trace. Panel (b), ``Verification approach richness,'' shows the number of participants that identified verification approaches. 14 within the Raw Trace and Space-Filling Nodes Condition; and 11 within the Sequential Timeline. Boxplots show a higher median count of identified approaches for both visualizations. Panel (c), ``Verification share estimation error,'' shows a lower median error for both visualizations compared to the Raw Trace. The difference is significant between the Raw Trace and Space-Filling Nodes.}
    \label{fig:rq1_panel}
\end{figure}

\subsubsection*{Strategy Summary Quality}
We found that participants wrote higher-quality summaries of the model's reasoning strategy when using the \retrace's visualizations. On our 3-point quality scale, the Space-Filling Nodes layout and the Sequential Timeline both yielded a higher median score ($Mdn{=}1.50$) than the \emph{Raw Trace} ($Mdn{=}1.00$; Fig.~\ref{fig:rq1_panel} (a)). Participants explained that \retrace's ability to provide a structured overview (~\ref{theme:1}), changed their comprehension strategy. P14 stated, the baseline is \textit{``a book without chapters where you have to read the whole text to know what is going on''}. In contrast, the visualization was \textit{``more like a book with chapters and sub-chapters [...] visually presented very nicely. So it was even easier than just having chapters.''} Their comprehension shifted to a non-linear approach, compared to the baseline (\ref{theme:3}. {``For the visualization, I can just choose which part I'm interested in and I can just go into it.''} (P10) This ability to get a high-level overview first and then selectively explore was a common sentiment among participants.

\subsubsection*{Verification Share and Approaches}
Participants using \retrace's visualizations identified more and described verification strategies more richly than with the Raw Trace. We evaluated if participants detected verification approaches, and if so how many qualitative ones. With both the Sequential Timeline ($N{=}11$, $Mdn{=}3.00$) and Space-Filling ($N{=}14$, $Mdn{=}3.00$) layouts resulting in higher median counts than the Raw Trace ($N{=}14$, $Mdn{=}2.00$; Fig.~\ref{fig:rq1_panel}b)). Additionally, estimation of the reasoning share spent on verification was also more accurate. Their median absolute error was lower for the Sequential Timeline ($Mdn{=}0.30pp$) and significantly lower for the Space-Filling Nodes ($p_{\text{Holm}}{=}0.001$; $Mdn{=}0.30pp$) compared to the Raw Trace($Mdn{=}14.60pp$; Fig.~\ref{fig:rq1_panel}c)). Within the visualizations, this task was straightforward because \retrace showed the share per phase directly. However, the results show that understanding and identifying distinct phases in the raw trace is a major challenge. Overall, \retrace helped to distinguish between different strategies that were otherwise blurred together in the monolithic text. The visualizations helped to \textit{``to pinpoint the main verification strategies that LLM had''} (P13). Participants could see the \textit{``the logic behind the steps that are being taken and how they connect to each other in a broader sense''} (P2).

\subsubsection*{Confidence and Redundancy Identification}
\retrace also improved participants' ability to pinpoint when the model became confident in its answer. Accuracy was highest for Space-Filling Nodes ($N{=}13$), followed by the Raw Trace ($N{=}11$), and was lowest for the Sequential Timeline ($N{=}8$). However, four participants within the visualizations pointed to the confidence step much earlier, inside the Initial Solution phase. Describing this insight into the reasoning as: \textit{``Actually, in all three tasks, [the LRM] did already solve the problem after the second phase. And if you just take that second phase in itself that contains all the logic, it's actually solving the problem''} (P1).
Redundancy was reported frequently across all conditions (Raw $N{=}15$; Space-Filling Nodes and Sequential Timeline $N{=}14$). However, exploring visualized traces (\ref{theme:5}), they were surprised to see \textit{``much time the LLM actually spent on re-evaluating''}, as P8 explained. While participants could simply read the raw steps in all conditions, contextualizing these from a local to a global understanding was hard in the Raw Trace. To grasp the model's broader pattern of confidence and redundant verifications, suggesting a difference in the participants' comprehension, requires the structural overview provided by \retrace.

\subsection{Does ReTrace reduce workload and task time and increase perceived ease? (RQ2)}
\begin{figure}[t]
    \centering
    \includegraphics[width=0.5\linewidth]{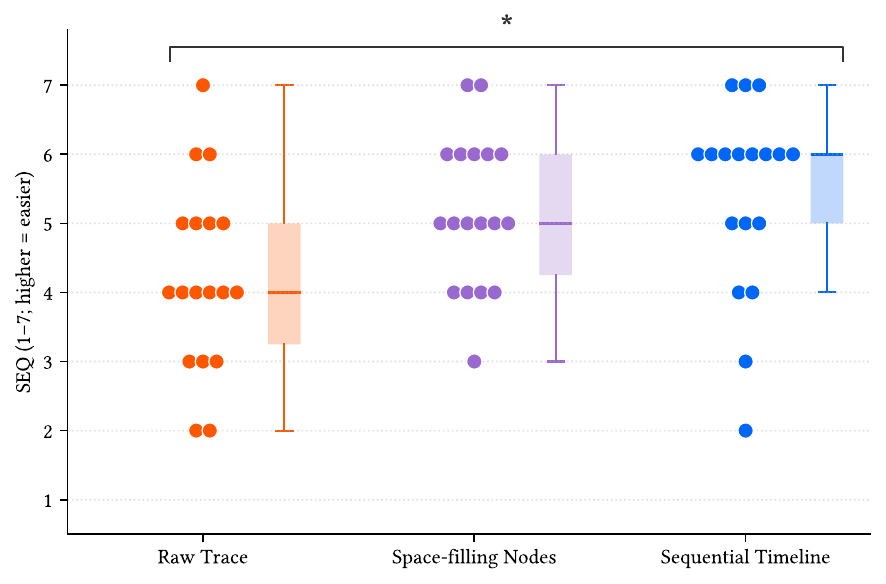}
    \caption{
    Perceived ease of use (RQ2) measured by the Single Ease Question (SEQ) on a 7-point scale (higher is easier). Each dot represents one participant ($N{=}18$). Boxplots show median and IQR; a star (*) indicates a statistically significant difference ($p_{\text{Holm}}<0.05$).
    }
    \Description{Three boxplots of perceived ease of use measured on a 7-point scale, where higher is easier. Additionally a scatterplot shows the answers of individual participants.  The chart compares the Raw Trace, Space-Filling Nodes, and Sequential Timeline conditions. The median rating is highest for the Sequential Timeline with six, followed by Space-Filling Nodes with five, and lowest for the Raw Trace with four. The difference between Sequential Timeline and Raw Trace is statistically significant.}
    \label{fig:seq-dot}
\end{figure}

\begin{figure}[t]
    \centering
    \includegraphics[width=0.95\linewidth]{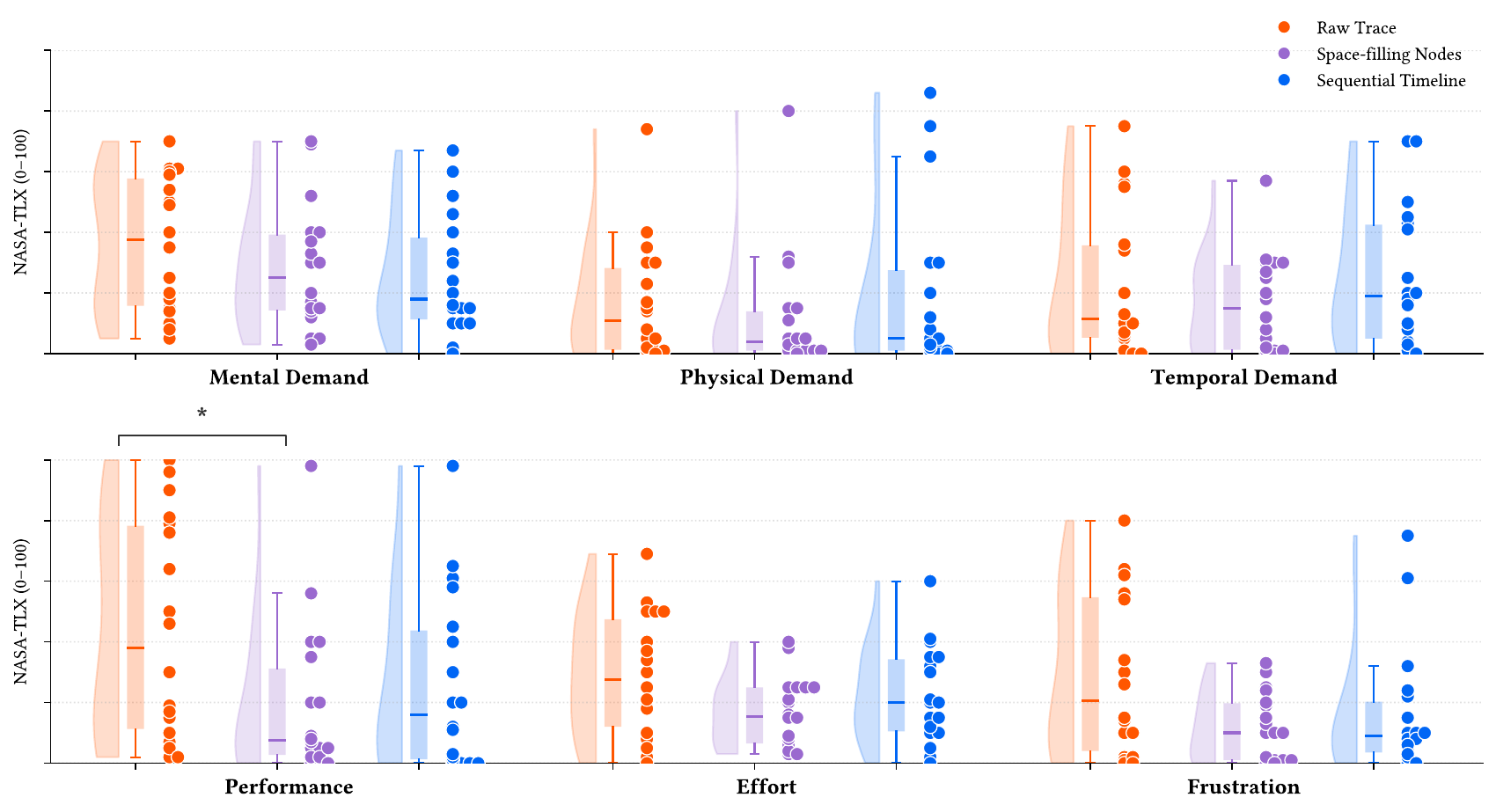}
    \caption{Perceived workload (RQ2) across six NASA-TLX dimensions (0-100 scale, lower is better). Boxplots show the median and IQR for each condition. A star (*) indicates a statistically significant difference ($p_{\text{Holm}}<0.05$) for the Performance dimension between Raw Trace and Space-Filling Nodes. 
    }
    \Description{Six plots showing perceived workload across the NASA-TLX dimensions on a 100-point scale. Each plot contains three boxplots to compare the Raw Trace, Space-Filling Nodes, and Sequential Timeline conditions. For dimensions where lower is better, such as Mental Demand and Frustration, the visualizations show lower median scores than the Raw Trace. Notably, on the Performance dimension, the Space-Filling Nodes condition shows a significantly higher lower score than the Raw Trace.}
    \label{fig:nasa-dot}
\end{figure}

\subsubsection*{Singe Ease Question and NASA\textendash TLX}
Participants perceived the trials as easier using \retrace. They rated the Sequential Timeline ($Mdn{=}6$) as easiest, followed by the Space-Filling Nodes ($Mdn{=}5$) and the Raw Trace($Mdn{=}4$) as most difficult, respectively. The difference between the first and last condition was significant ($p_{\text{Holm}}{=}0.016$; Fig.~\ref{fig:seq-dot}). In line with these ratings, perceived workload was also lower with \retrace. The difference was significant between the Space-Filling Nodes ($Mdn{=}12.50$) and Raw Trace ($p_{\text{Holm}}{=}0.014$; $Mdn{=}20.50$), and a low trend in the Sequential Timeline ($Mdn{=}15.00$). Inside the dimension of Performance, the rating of success in accomplishing the task was also significantly better between Space-Filling Nodes and the Raw Trace ($p_{\text{Holm}}{=}0.003$; Fig.~\ref{fig:nasa-dot}).
Participants explained these improvements by pointing to the structure replacing exhaustive navigation. The baseline felt like \textit{“mainly scrolling endlessly and going up and down constantly”} (P12). In contrast, the visualizations made it easier by taking \textit{“some of the mental work away by reducing the labor on formatting in your brain”} (P6). The Sequential Timeline was often \textit{“the least amount of physical movement, like clicks and everything”} (P15), though in dense traces it could be \textit{“hard to read and navigate through”} (P4). Overall, we observed a common theme in cognitive offloading (\ref{theme:2}), where \retrace automates tasks that would otherwise require manual work. Concretely, this offloading appeared in two areas: reduced memorization and reduced manual structuring, while physical and temporal demand were rated similarly across conditions.

\subsubsection*{Task Time}
\begin{figure}[h]
    \centering
    \includegraphics[width=0.5\linewidth]{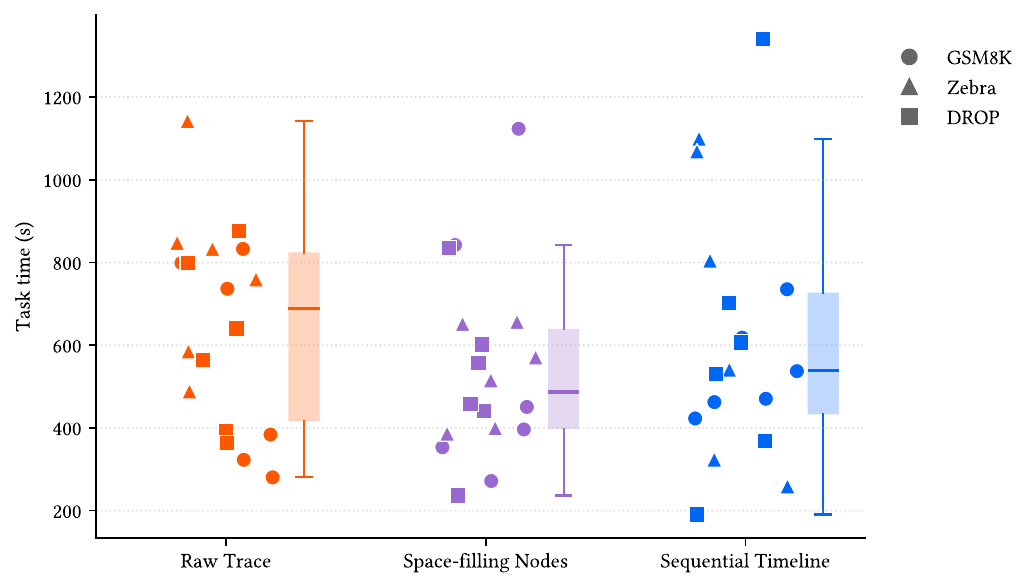}
    \caption{
    Task completion time in seconds (lower is faster). Each point represents a single trial, with marker shape indicating the dataset. Boxplots show median and IQR for each condition. No significant differences were found.
    }
    \Description{Boxplots of task completion time in seconds across the three conditions. The chart shows that median task times and interquartile ranges are similar across the Raw Trace, Space-Filling Nodes, and Sequential Timeline conditions, with no significant differences found. Each data point in the scatterplot represents a single trial, and its shape indicates the dataset used for that trial.}
    \label{fig:task-time}
\end{figure}

We found no significant differences in task completion time between the conditions. However, visual analysis suggests slightly faster times with \retrace's visualizations (Fig.~\ref{fig:task-time}). 
During the study, we observed participants often exploring the interface first before starting the tasks. Despite training examples, several participants spent time thoroughly exploring the visualizations. This novelty effect may explain the non-significant differences. An alternative explanation is that interaction overhead in the visualizations (clicking, expanding, navigating) may offset benefits for speed.

\subsection{How Does ReTrace Support Calibrated Trust and Confidence? (RQ3)}
\subsubsection*{Correctness Judgment and Confidence Calibration}
Accuracy was high across all conditions, which we attribute to easy, closed-ended items with clear ground truth that were straightforward to verify (Tab.~\ref{tab:materials}). Raw Trace achieved 94.4\% (17/18), Sequential Timeline 88.9\% (16/18), and Space-Filling Nodes 94.4\% (17/18). Confidence levels were similar and clustered at 4–5 (Raw Trace $Mdn{=}5.00$, Sequential Timeline $Mdn{=}4.00$, Space-Filling Nodes $Mdn{=}5.00$). Some participants reported more doubt after seeing repeated self-checks and hesitations in the reasoning. Most others described the opposite pattern and linked trust to better comprehension. P5 explained that with the Space-Filling Nodes they were \textit{“the most certain that [they] didn’t make a mistake in my understanding.”} Other rationals were that the organization made it \textit{“easier to mentally organize the info” and that they “made me more confident in my answers.”} (P15). P13 emphasized that confidence \textit{“comes from understanding how it got this result.”}

\subsubsection*{Perceived Logical Soundness}
\begin{figure}[t]
    \centering
    \includegraphics[width=0.5\linewidth]{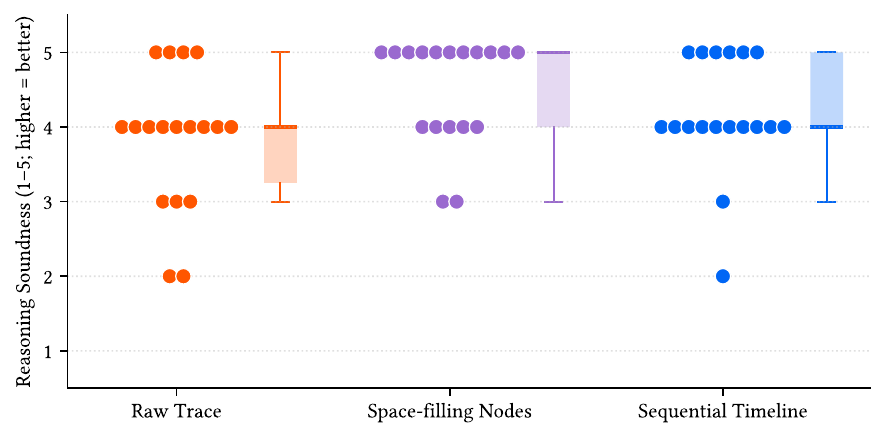}
    \caption{
    Ratings of perceived logical soundness (RQ3) on a 5-point scale (higher is better). Each dot represents a single trial. Boxplots show median and IQR for each condition.
    }
    \Description{Boxplot of perceived logical soundness ratings on a 5-point scale, where a higher score is better. The chart compares the Raw Trace, Space-Filling Nodes, and Sequential Timeline conditions. Additional scatter dots show individual responses from participants. The median rating is highest for Space-Filling Nodes at 5, while the median for both the Timeline and Raw Trace is 4.}
    \label{fig:reasoning-soundness}
\end{figure}

Similarly, \retrace showed slightly better results when users rated the logical soundness of the reasoning itself. The Space-Filling Nodes were rated highest ($Mdn{=}5.00$), followed by the Timeline ($Mdn{=}4.00$), and the Raw Trace ($Mdn{=}4.00$; Fig.~\ref{fig:reasoning-soundness}). However, the structural overview (\ref{theme:1}) provided by \retrace led to a better assessment of the model's inner working and thereby to a better understanding of the internal structure of reasoning. P3 noted that the visualizations \textit{``made me find out that there is a part that is doing the verification, whereas the raw trace didn't tell me that.''} With \retrace's visualization participants could see \textit{``logic behind the steps''} (P2).

\subsection{Overall Usefulness and User Preference}
\begin{figure}[t]
    \centering
    \includegraphics[width=0.5\linewidth]{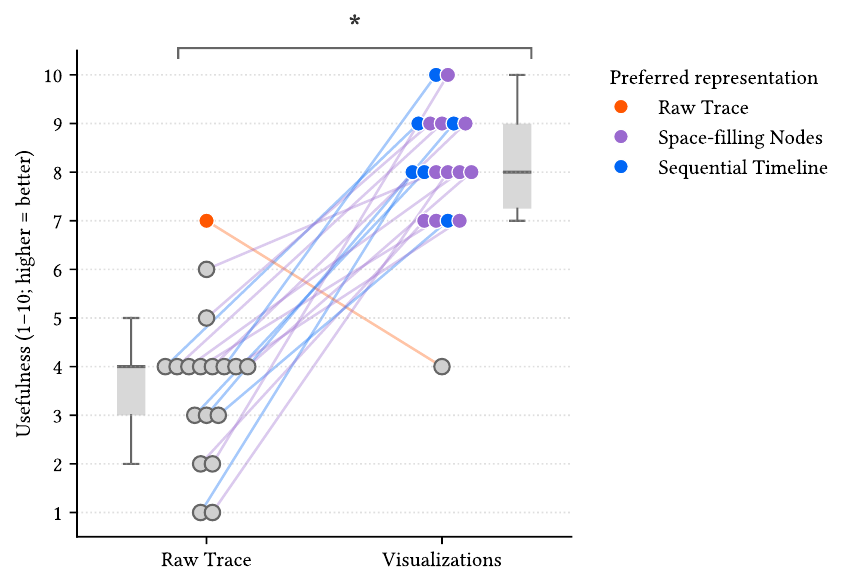}
    \caption{
    Overall usefulness ratings and final user preference. The paired slopegraph shows each participant's usefulness rating (1-10 scale) for the Raw Trace (left) versus the \retrace visualizations (right). The color of each participant's line and data points corresponds to their single, final preferred representation, as indicated in the legend. Eleven participants preferred the Space-Filling Nodes, six the Sequential Timeline and one the Raw Trace.  Boxplots show the median and IQR for the ratings. A star (*) indicates a statistically significant difference in usefulness ($p_{\text{Holm}}<0.05$).
    }
    \Description{Paired slopegraph showing participants' overall usefulness ratings on a 10-point scale. The chart compares ratings for the Raw Trace on the left against the ReTrace visualizations on the right. Nearly all lines, each representing one participant, slope upward, indicating a strong and statistically significant preference for the visualizations over the Raw Trace. The color of the lines and the dot with the higher usefulness rating, indicates which visualization was preferred. Eleven participants preferred the Space-Filling Nodes, six the Sequential Timeline and one the Raw Trace.}
    \label{fig:usefulness-preference}
\end{figure}

According to our previous results, \retrace also achieves high scores in the usefulness rating. We find that a significant majority preferred the reasoning visualizations ($Mdn{=}8$) over the raw reasoning ($p_{\text{Holm}}{=}0.0003$, $Mdn{=}4$). Almost all participants ranked either Space-Filling Nodes ($N{=}11$) or Timeline Nodes ($N{=}6$) as their preferred reasoning representation. Only one participant rated the Raw Trace as more usefulness and preferred the Raw Trace ($N{=}1$; Fig.~\ref{fig:usefulness-preference}). The answers in the free text response can trace back to the interview themes discussed previously. Participants explained that the provided detail on demand overview (\ref{theme:1}) made it \textit{``much easier to understand a small amount of large ideas first and then disentangle them level by level''} (13). While the reduced mental labor while exploring (\ref{theme:2}) with \retrace helped to counter the loss of \textit{``valuable insights from LLM trace can be missed because I just give up on reading it''} (P13). Overall, we found a preference for our Space-Filling layout (\ref{theme:6}).

\begin{table}[th]
    \caption{Summary of qualitative participant feedback, organized by themes (T1-T6).
    }
    \Description{Table summarizing six themes from the qualitative analysis of participant feedback. The themes listed include 'Providing a Structured Overview', 'Reducing Cognitive Load', 'Enhancing Navigation and Exploration', 'Increasing Confidence in Comprehension', 'Revealing Flaws in the LLM's Reasoning' and 'Preference for Space-Filling Layout'. For each theme, the table provides a brief description, representative quotes from participants, and a list of the participant IDs who contributed comments related to that theme. For example Theme 'Providing a Structured Overview (T1)' is described as 'Participants found ReTrace valuable the visualizations structured reasoning process overview'. The example shows ``Overall, I would say all visualizations, due to the summaries and the structure, made it easier to get an overview.'' and Participants 'P1, P3, P8, P10, P12, P14–P18'.}
    \label{tab:interview-findings}
\centering
\small
\setlength{\tabcolsep}{4pt}
\renewcommand{\arraystretch}{1.0}
\begin{tabularx}{\linewidth}{
  >{\raggedright\arraybackslash}p{0.16\linewidth}
  >{\raggedright\arraybackslash}p{0.25\linewidth}
  >{\raggedright\arraybackslash}X %
  >{\raggedright\arraybackslash}p{0.08\linewidth}}
\toprule
\textbf{Theme} & \textbf{Description} & \textbf{Representative Examples} & \textbf{Participants} \\
\midrule
Providing a Structured Overview \newline(\themelabel{theme:1}) 
& 
Participants found \retrace valuable the visualizations structured reasoning process overview.
&
\textit{``Overall, I would say all visualizations, due to the summaries and the structure, made it easier to get an overview.'' \newline
``...the raw thing was like without chapters and without structure and the visualization was like more with the book with chapters and sub chapters.''}
& 
P1, P3, P8, P10, P12, P14–P18
\\
\midrule
Reducing Cognitive Load \newline(\themelabel{theme:2}) 
& 
Participants found the visualizations reduced the cognitive load of understanding the reasoning compared to the raw text.
& 
\textit{``It's much less overwhelming. And then you can, as you wish, explore each of the ideas by opening the levels below.'' \newline
``...the visualization takes some of the mental work away by reducing the labor on formatting in your brain... It did a little bit of that pre-work for you.''}
& 
P1–P4, P6, P10, P11, P14 
\\
\midrule
Enhancing Navigation and Exploration \newline(\themelabel{theme:3})
& 
Participants valued the interactive features of the visualizations for non-linear exploration and drilling down into details.
& 
\textit{``I couldn't do in the plain text version was seeing the logic behind the steps... how they connect to each other in a broader sense''\newline
``...with the visualization, you could leave open a particular step... and traverse through... And at the same time, you do not lose the track.''}
& 
P1–P3, P5, P8, P10, P12–P15
\\
\midrule
Increasing Confidence in Comprehension \newline(\themelabel{theme:4})
& 
The structured visualizations increased participants' confidence in their ability to comprehend the model's reasoning.
& 
\textit{``Yes, it helped me to become more confident because I could see where the LNM made its specific reasonings.''\newline
``It was easier to mentally organize the info and made me more confident in my answers.''}
& 
P2, P5, P6, P10, P12–P15, P17
\\
\midrule
Revealing Flaws in the LLM's Reasoning \newline(\themelabel{theme:5})
& 
Participants noted how the visualizations clearly exposed inefficiencies and redundancies in the LRM's reasoning process.
& 
\textit{``I didn't knew how much time the LLM actually spent on re-evaluating so this was quite interesting for me.'' \newline
``...many tasks repeat themselves. And he [the LRM] thinks again, thinks again, thinks again.''}
& 
P1, P2, P6, P8, P11–P13
\\
\midrule
Preference for Space-Filling Layout \newline(\themelabel{theme:6})
&
Participants expressed a clear preference for the Space-Filling layout, finding it more structured and less cluttered than the timeline view.
&
\textit{``The last diagram with the rectangles... It's the best way to see the hierarchy and that was the most useful diagram for me.''\newline
``I felt like those [Space-Filling nodes] were the easiest to process visually, because it was more structured than the timeline.''}
&
P4, P5, P7, P8, P10, P11, P13, P17, P18
\\
\bottomrule
\end{tabularx}

\end{table}

\section{Discussion}
Our work explored how interactive visualizations of reasoning traces can support human comprehension and sensemaking. The findings from our user study highlight the promising benefits and opportunities of \retrace. For almost all of our evaluation questions, perceived ease, and workload, our reasoning visualizations achieve better results compared to the baseline. Below, we discuss implications and lessons for LLM providers, designers, and end users, as well as limitations and future work. 

\subsection{Design Implications for LLM Providers and Designers} 
Current LRM interfaces often dump or hide reasoning traces, treating the internal reasoning as a by-product rather than a first-class feature. Our results point in a different direction. Structured, visual explanations can support effective human–AI interaction, aligning with established guidelines for building AI systems that make clear why they did what they did~\cite{amershiGuidelinesHumanAIInteraction2019}.

\subsubsection*{Visualize the Process, Reduce the ``Black Box''}
\retrace converts a verbose, linear trace into a structured, explorable view. Participants shifted from passive reading to active exploration of the model's ``thought process.'' In line with the goals of human-centered XAI, seeking to empower users to probe and understand a system's reasoning rather than passively accept its output~\cite{liaoQuestioningAIInforming2020}. For participants, the raw reasoning was \textit{``a book without chapters,”} and with visualized support, the trace became \textit{``a book with chapters''} (P14). Our visualizations helped them better understand internal thinking by supporting them with the mentally demanding, yet easily automated, task of grouping and summarizing. Thereby giving them insights into the model's logic~\cite{doshivelezRigorousScienceInterpretable2017}. These steps directly reduce the ``black box'' nature of AI reasoning by visualizing the ``latent variables in the model's computation''~\cite{korbakChainThoughtMonitorability2025}.

\subsubsection*{Enable Critical Assessment by Revealing Flaws}
Visualizations helped participants not only to see the model's strategy but also its inefficiencies. They noticed loops, redundant checks, and long re-evaluation (\ref{theme:5}). Without them, they just \textit{“didn’t know how much time the LLM actually spent on re-evaluating.”} (P8). Exposing these imperfections is a feature, not a weakness. Reasoning visualizations calibrate trust by revealing the real, iterative path to an answer, supporting appropriate reliance and mitigating overtrust~\cite{hoffTrustAutomationIntegrating2015}. Especially important given that system failures often have a disproportionately stronger negative impact on trust than successes have a positive one~\cite{yuUserTrustDynamics2017}. It allows users to see the model not as a perfect oracle, but as a computational tool with a discernible, sometimes flawed, process. Thus, fosters a more mature and critical mindset, which is essential for responsible AI deployment.

\subsection{Lessons for End-Users}
Our results suggest how end-users can work with LRMs. Moving from passive reading to an active, critical partnership.

\subsubsection*{Learning to ``Read'' the Reasoning Trace}
With the right scaffolds, users learn to interpret the reasoning process. \retrace helped our participants to see strategies, points of confidence, and uncertainty. This process moves beyond simple verification and fosters a deeper form of AI literacy~\cite{longAILiteracyFinding2023}, teaching users how to critically engage with a model's output. Trust “\textit{comes from understanding how it got this result}” (P13). Systems like \retrace can act as training wheels. They help users build habits for evaluating machine reasoning. This echoes the self-explanation effect, where actively explaining a process yields more robust understanding than passively receiving an answer~\cite{chi1994eliciting, wenEnhancingSelfExplanationStudent2025}. Supporting user-driven sensemaking in human-centered XAI~\cite{mueller2021principles} and learning from the reasoning trace itself.

\subsubsection*{From Answer-Checking to Process Auditing}
Participants changed their approach. With the raw traces, they searched for an answer and verified it in a demanding process. With \retrace, they could audit the entire reasoning process. They distinguished strategies, identified redundant steps, and assessed overall soundness. As P1 stated, \textit{“what I expressly could do in the visualization that I couldn't do in the plain text version was seeing the logic behind the steps that are being taken and how they connect to each other in a broader sense.”} This transition restores human agency and control when interacting with opaque AI systems~\cite{shneidermanHumanCenteredAI2022}. \retrace makes the reasoning process itself accessible and thereby can support a stronger feeling of controllability and justified confidence.

\subsection{Limitations \& Future Work}
Our study shows that visualizing reasoning traces is useful, but the scope has several limitations. \retrace targets only single textual traces from DeepSeek-R1 because, at the time of the study, it was one of the very few open-weights LRMs that exposed a whole reasoning trace. %
We rely on an external LLM for grouping and summarization, which introduces potential interpretation errors. We mitigated the risk by adopting a validated taxonomy and linking every group back to verbatim steps for verification. 
Our study setup also limits ecological validity. All participants worked with static, pre-generated traces and a sample leaning toward computer science and engineering backgrounds. Additionally, we relied on a pre-defined set of tasks to assess the reasoning trace, which may differ from tasks and challenges in real-world workflows.

Our work presents multiple opportunities for future work. We envision a shift from post-hoc inspection to real-time visualization. Thereby, users could monitor strategy, catch unproductive paths early, and steer behavior with targeted prompts or controls. Additionally, future work could further improve the detection of reasoning loops, redundant verification, and low-value branches, and collapse them into concise annotations. Promising directions also include expanding our approach beyond linear monologues. Agentic systems with tools, parallel branches, and external context may require richer representations—hybrid trees and graphs that expose alternatives, abandoned paths, and data flow.

\section{Conclusion}
In this work, we presented \retrace, an interactive visualization system designed to address the sensemaking challenges posed by verbose LRM traces. \retrace transforms unstructured textual reasoning trace into two distinct visualization layouts: a hierarchical view for structural analysis and a sequential Timeline for chronological understanding. Our user study demonstrates that abstracting raw text into these interactive visualizations improves participant comprehension and reduces perceived workload. From our findings, we draw three design implications for future systems in this space: the need to support abstraction and semantic chunking to offload cognitive effort; the importance of visualizing the iterative process of reasoning beyond its literal path; and balance chronology with hierarchy. Recognizing the urgency of human-centered AI explainability, we hope our work inspires future research on interactive visualization approaches for making complex AI processes transparent and comprehensible.

\bibliographystyle{ACM-Reference-Format}
\bibliography{base}

@misc{comaniciGemini25Pushing2025,
  title = {Gemini 2.5: {{Pushing}} the {{Frontier}} with {{Advanced Reasoning}}, {{Multimodality}}, {{Long Context}}, and {{Next Generation Agentic Capabilities}}},
  shorttitle = {Gemini 2.5},
  author = {Comanici, Gheorghe and Bieber, Eric and Schaekermann, Mike and Pasupat, Ice and Sachdeva, Noveen and Dhillon, Inderjit and Blistein, Marcel and Ram, Ori and Zhang, Dan and others},
  year = {2025},
  month = jul,
  number = {arXiv:2507.06261},
  eprint = {2507.06261},
  primaryclass = {cs},
  publisher = {arXiv},
  doi = {10.48550/arXiv.2507.06261},
  urldate = {2025-07-11},
  archiveprefix = {arXiv},
  langid = {english}
}

@misc{marjanovicDeepSeekR1ThoughtologyLets2025,
  title = {{{DeepSeek-R1 Thoughtology}}: {{Let}}'s {$<$}think{$>$} about {{LLM Reasoning}}},
  shorttitle = {{{DeepSeek-R1 Thoughtology}}},
  author = {Marjanovi{\'c}, Sara Vera and Patel, Arkil and Adlakha, Vaibhav and Aghajohari, Milad and BehnamGhader, Parishad and Bhatia, Mehar and Khandelwal, Aditi and Kraft, Austin and Krojer, Benno and L{\`u}, Xing Han and Meade, Nicholas and Shin, Dongchan and Kazemnejad, Amirhossein and Kamath, Gaurav and Mosbach, Marius and Sta{\'n}czak, Karolina and Reddy, Siva},
  year = {2025},
  month = apr,
  number = {arXiv:2504.07128},
  eprint = {2504.07128},
  primaryclass = {cs},
  publisher = {arXiv},
  doi = {10.48550/arXiv.2504.07128},
  urldate = {2025-05-05},
  archiveprefix = {arXiv},
  langid = {english},
}

@article{schmidt2008sankey,
  title={The Sankey diagram in energy and material flow management: part II: methodology and current applications},
  author={Schmidt, Mario},
  journal={Journal of industrial ecology},
  volume={12},
  number={2},
  pages={173--185},
  year={2008},
  publisher={Wiley Online Library}
}

@article{ferrante1987program,
  title={The program dependence graph and its use in optimization},
  author={Ferrante, Jeanne and Ottenstein, Karl J and Warren, Joe D},
  journal={ACM Transactions on Programming Languages and Systems (TOPLAS)},
  volume={9},
  number={3},
  pages={319--349},
  year={1987},
  publisher={ACM New York, NY, USA}
}

@article{devkota2020ccnav,
  title={CcNav: Understanding compiler optimizations in binary code},
  author={Devkota, Sabin and Aschwanden, Pascal and Kunen, Adam and Legendre, Matthew and Isaacs, Katherine E},
  journal={IEEE transactions on visualization and computer graphics},
  volume={27},
  number={2},
  pages={667--677},
  year={2020},
  publisher={IEEE}
}

@article{balmas2004displaying,
  title={Displaying dependence graphs: a hierarchical approach},
  author={Balmas, Francoise},
  journal={Journal of Software Maintenance and Evolution: Research and Practice},
  volume={16},
  number={3},
  pages={151--185},
  year={2004},
  publisher={Wiley Online Library}
}

@inproceedings{toprak2014lightweight,
  title={Lightweight structured visualization of assembler control flow based on regular expressions},
  author={Toprak, Sibel and Wichmann, Arne and Schupp, Sibylle},
  booktitle={2014 Second IEEE Working Conference on Software Visualization},
  pages={97--106},
  year={2014},
  organization={IEEE}
}

@inproceedings{dingReVISitSupportingScalable2023,
  title = {{{reVISit}}: {{Supporting Scalable Evaluation}} of {{Interactive Visualizations}}},
  shorttitle = {{{reVISit}}},
  booktitle = {2023 {{IEEE Visualization}} and {{Visual Analytics}} ({{VIS}})},
  author = {Ding, Yiren and Wilburn, Jack and Shrestha, Hilson and Ndlovu, Akim and Gadhave, Kiran and Nobre, Carolina and Lex, Alexander and Harrison, Lane},
  year = {2023},
  month = oct,
  pages = {31--35},
  publisher = {IEEE},
  address = {Melbourne, Australia},
  doi = {10.1109/VIS54172.2023.00015},
  urldate = {2025-08-20},
  copyright = {https://doi.org/10.15223/policy-029},
  isbn = {979-8-3503-2557-7},
  langid = {english}
}

@misc{korbakChainThoughtMonitorability2025,
  title = {Chain of {{Thought Monitorability}}: {{A New}} and {{Fragile Opportunity}} for {{AI Safety}}},
  shorttitle = {Chain of {{Thought Monitorability}}},
  author = {Korbak, Tomek and Balesni, Mikita and Barnes, Elizabeth and Bengio, Yoshua and Benton, Joe and Bloom, Joseph and Chen, Mark and Cooney, Alan and Dafoe, Allan and Dragan, Anca and Emmons, Scott and Evans, Owain and Farhi, David and Greenblatt, Ryan and Hendrycks, Dan and Hobbhahn, Marius and Hubinger, Evan and Irving, Geoffrey and Jenner, Erik and Kokotajlo, Daniel and Krakovna, Victoria and Legg, Shane and Lindner, David and Luan, David and M{\k a}dry, Aleksander and Michael, Julian and Nanda, Neel and Orr, Dave and Pachocki, Jakub and Perez, Ethan and Phuong, Mary and Roger, Fabien and Saxe, Joshua and Shlegeris, Buck and Soto, Mart{\'i}n and Steinberger, Eric and Wang, Jasmine and Zaremba, Wojciech and Baker, Bowen and Shah, Rohin and Mikulik, Vlad},
  year = {2025},
  month = jul,
  number = {arXiv:2507.11473},
  eprint = {2507.11473},
  primaryclass = {cs},
  publisher = {arXiv},
  doi = {10.48550/arXiv.2507.11473},
  urldate = {2025-07-30},
  archiveprefix = {arXiv},
  langid = {english}
}

@misc{deepseek-aiDeepSeekR1IncentivizingReasoning2025a,
  title = {{{DeepSeek-R1}}: {{Incentivizing Reasoning Capability}} in {{LLMs}} via {{Reinforcement Learning}}},
  shorttitle = {{{DeepSeek-R1}}},
  author = {{DeepSeek-AI} and Guo, Daya and others},
  year = {2025},
  month = jan,
  number = {arXiv:2501.12948},
  eprint = {2501.12948},
  primaryclass = {cs},
  publisher = {arXiv},
  doi = {10.48550/arXiv.2501.12948},
  urldate = {2025-08-24},
  archiveprefix = {arXiv},
  langid = {english}
}

@misc{liReasonGraphVisualisationReasoning2025,
  title = {{{ReasonGraph}}: {{Visualisation}} of {{Reasoning Paths}}},
  shorttitle = {{{ReasonGraph}}},
  author = {Li, Zongqian and Shareghi, Ehsan and Collier, Nigel},
  year = {2025},
  month = mar,
  number = {arXiv:2503.03979},
  eprint = {2503.03979},
  primaryclass = {cs},
  publisher = {arXiv},
  doi = {10.48550/arXiv.2503.03979},
  urldate = {2025-03-20},
  archiveprefix = {arXiv}
}

@misc{vigVisualizingAttentionTransformerBased2019,
  title = {Visualizing {{Attention}} in {{Transformer-Based Language Representation Models}}},
  author = {Vig, Jesse},
  year = {2019},
  month = apr,
  number = {arXiv:1904.02679},
  eprint = {1904.02679},
  primaryclass = {cs},
  publisher = {arXiv},
  doi = {10.48550/arXiv.1904.02679},
  urldate = {2025-08-30},
  archiveprefix = {arXiv},
  langid = {english}
}

@misc{pierseTransformersInterpret2021,
  title = {Transformers {{Interpret}}},
  author = {Pierse, Charles},
  year = {2021},
  month = feb,
  urldate = {2025-08-30},
  copyright = {Apache-2.0}
}

@misc{healyFrames_of_mindAnimatingR1s,
  title = {Frames\_of\_mind: {{Animating R1}}'s Thoughts.},
  author = {Healy, Dan},
  urldate = {2025-08-30},
  howpublished = {https://github.com/dhealy05/frames\_of\_mind}
}

@misc{dibiaLIDAToolAutomatic2023,
  title = {{{LIDA}}: {{A Tool}} for {{Automatic Generation}} of {{Grammar-Agnostic Visualizations}} and {{Infographics}} Using {{Large Language Models}}},
  shorttitle = {{{LIDA}}},
  author = {Dibia, Victor},
  year = {2023},
  month = jun,
  number = {arXiv:2303.02927},
  eprint = {2303.02927},
  primaryclass = {cs},
  publisher = {arXiv},
  doi = {10.48550/arXiv.2303.02927},
  urldate = {2025-08-30},
  archiveprefix = {arXiv},
  langid = {english}
}

@misc{geroSupportingSensemakingLarge2024,
  title = {Supporting {{Sensemaking}} of {{Large Language Model Outputs}} at {{Scale}}},
  author = {Gero, Katy Ilonka and Swoopes, Chelse and Gu, Ziwei and Kummerfeld, Jonathan K. and Glassman, Elena L.},
  year = {2024},
  month = jan,
  number = {arXiv:2401.13726},
  eprint = {2401.13726},
  primaryclass = {cs},
  publisher = {arXiv},
  doi = {10.48550/arXiv.2401.13726},
  urldate = {2025-08-30},
  archiveprefix = {arXiv}
}

@inproceedings{arawjoChainForgeVisualToolkit2024,
  title = {{{ChainForge}}: {{A Visual Toolkit}} for {{Prompt Engineering}} and {{LLM Hypothesis Testing}}},
  shorttitle = {{{ChainForge}}},
  booktitle = {Proceedings of the {{CHI Conference}} on {{Human Factors}} in {{Computing Systems}}},
  author = {Arawjo, Ian and Swoopes, Chelse and Vaithilingam, Priyan and Wattenberg, Martin and Glassman, Elena},
  year = {2024},
  month = may,
  eprint = {2309.09128},
  primaryclass = {cs},
  pages = {1--18},
  doi = {10.1145/3613904.3642016},
  urldate = {2025-08-30},
  archiveprefix = {arXiv}
}

@inproceedings{jiangGraphologueExploringLarge2023a,
  title = {Graphologue: {{Exploring Large Language Model Responses}} with {{Interactive Diagrams}}},
  shorttitle = {Graphologue},
  booktitle = {Proceedings of the 36th {{Annual ACM Symposium}} on {{User Interface Software}} and {{Technology}}},
  author = {Jiang, Peiling and Rayan, Jude and Dow, Steven P. and Xia, Haijun},
  date = {2023-10-29},
  pages = {1--20},
  publisher = {ACM},
  location = {San Francisco CA USA},
  doi = {10.1145/3586183.3606737},
  url = {https://dl.acm.org/doi/10.1145/3586183.3606737},
  urldate = {2025-08-31},
  eventtitle = {{{UIST}} '23: {{The}} 36th {{Annual ACM Symposium}} on {{User Interface Software}} and {{Technology}}},
  isbn = {979-8-4007-0132-0},
  langid = {english},
}

@online{huangReasoningLargeLanguage2023,
  title = {Towards {{Reasoning}} in {{Large Language Models}}: {{A Survey}}},
  shorttitle = {Towards {{Reasoning}} in {{Large Language Models}}},
  author = {Huang, Jie and Chang, Kevin Chen-Chuan},
  date = {2023-05-26},
  eprint = {2212.10403},
  eprinttype = {arXiv},
  eprintclass = {cs},
  doi = {10.48550/arXiv.2212.10403},
  url = {http://arxiv.org/abs/2212.10403},
  urldate = {2025-08-31},
  langid = {english},
  pubstate = {prepublished},
}

@online{yuNaturalLanguageReasoning2023,
  title = {Natural {{Language Reasoning}}, {{A Survey}}},
  author = {Yu, Fei and Zhang, Hongbo and Tiwari, Prayag and Wang, Benyou},
  date = {2023-05-13},
  eprint = {2303.14725},
  eprinttype = {arXiv},
  eprintclass = {cs},
  doi = {10.48550/arXiv.2303.14725},
  url = {http://arxiv.org/abs/2303.14725},
  urldate = {2025-08-31},
  langid = {english},
  pubstate = {prepublished},
}

@online{weiChainofThoughtPromptingElicits2023a,
  title = {Chain-of-{{Thought Prompting Elicits Reasoning}} in {{Large Language Models}}},
  author = {Wei, Jason and Wang, Xuezhi and Schuurmans, Dale and Bosma, Maarten and Ichter, Brian and Xia, Fei and Chi, Ed and Le, Quoc and Zhou, Denny},
  date = {2023-01-10},
  eprint = {2201.11903},
  eprinttype = {arXiv},
  eprintclass = {cs},
  doi = {10.48550/arXiv.2201.11903},
  url = {http://arxiv.org/abs/2201.11903},
  urldate = {2025-08-31},
  langid = {english},
  pubstate = {prepublished},
}

@online{kojimaLargeLanguageModels2023,
  title = {Large {{Language Models}} Are {{Zero-Shot Reasoners}}},
  author = {Kojima, Takeshi and Gu, Shixiang Shane and Reid, Machel and Matsuo, Yutaka and Iwasawa, Yusuke},
  date = {2023-01-29},
  eprint = {2205.11916},
  eprinttype = {arXiv},
  eprintclass = {cs},
  doi = {10.48550/arXiv.2205.11916},
  url = {http://arxiv.org/abs/2205.11916},
  urldate = {2025-08-31},
  langid = {english},
  pubstate = {prepublished}
}

@online{reynoldsPromptProgrammingLarge2021,
  title = {Prompt {{Programming}} for {{Large Language Models}}: {{Beyond}} the {{Few-Shot Paradigm}}},
  shorttitle = {Prompt {{Programming}} for {{Large Language Models}}},
  author = {Reynolds, Laria and McDonell, Kyle},
  date = {2021-02-15},
  eprint = {2102.07350},
  eprinttype = {arXiv},
  eprintclass = {cs},
  doi = {10.48550/arXiv.2102.07350},
  url = {http://arxiv.org/abs/2102.07350},
  urldate = {2025-09-02},
  langid = {english},
  pubstate = {prepublished}
}

@online{zelikmanSTaRBootstrappingReasoning2022,
  title = {{{STaR}}: {{Bootstrapping Reasoning With Reasoning}}},
  shorttitle = {{{STaR}}},
  author = {Zelikman, Eric and Wu, Yuhuai and Mu, Jesse and Goodman, Noah D.},
  date = {2022-05-20},
  eprint = {2203.14465},
  eprinttype = {arXiv},
  eprintclass = {cs},
  doi = {10.48550/arXiv.2203.14465},
  url = {http://arxiv.org/abs/2203.14465},
  urldate = {2025-08-31},
  langid = {english},
  pubstate = {prepublished}
}

@book{card1999readings,
  title={Readings in Information Visualization: Using Vision to Think.},
  author={Card, Stuart K and Mackinlay, Jock and Shneiderman, Ben},
  year={1999},
  publisher={Morgan Kaufmann Publishers Inc. San Francisco, CA, USA.}
}

@book{norman1986User,
author = {Norman, Donald A. and Draper, Stephen W.},
title = {User Centered System Design; New Perspectives on Human-Computer Interaction},
year = {1986},
isbn = {0898597811},
publisher = {L. Erlbaum Associates Inc.},
address = {USA}
}

@inproceedings{suhSensecapeEnablingMultilevel2023,
  title = {Sensecape: {{Enabling Multilevel Exploration}} and {{Sensemaking}} with {{Large Language Models}}},
  shorttitle = {Sensecape},
  booktitle = {Proceedings of the 36th {{Annual ACM Symposium}} on {{User Interface Software}} and {{Technology}}},
  author = {Suh, Sangho and Min, Bryan and Palani, Srishti and Xia, Haijun},
  date = {2023-10-29},
  pages = {1--18},
  publisher = {ACM},
  location = {San Francisco CA USA},
  doi = {10.1145/3586183.3606756},
  url = {https://dl.acm.org/doi/10.1145/3586183.3606756},
  urldate = {2025-08-31},
  eventtitle = {{{UIST}} '23: {{The}} 36th {{Annual ACM Symposium}} on {{User Interface Software}} and {{Technology}}},
  isbn = {979-8-4007-0132-0},
  langid = {english}
}

@inproceedings{shneiderman2003eyes,
    author = { Shneiderman, Ben },
    booktitle = { Visual Languages, IEEE Symposium on },
    title = {{ The Eyes Have It: A Task by Data Type Taxonomy for Information Visualizations }},
    year = {1996},
    volume = {},
    ISSN = {1049-2615},
    pages = {336},
    doi = {10.1109/VL.1996.545307},
    url = {https://doi.ieeecomputersociety.org/10.1109/VL.1996.545307},
    publisher = {IEEE Computer Society},
    address = {Los Alamitos, CA, USA},
    month =sep
}

@article{kochVarifocalReaderInDepthVisual2014,
  title = {{{VarifocalReader}} — {{In-Depth Visual Analysis}} of {{Large Text Documents}}},
  author = {Koch, Steffen and John, Markus and Worner, Michael and Muller, Andreas and Ertl, Thomas},
  date = {2014-12-31},
  journaltitle = {IEEE Transactions on Visualization and Computer Graphics},
  shortjournal = {IEEE Trans. Visual. Comput. Graphics},
  volume = {20},
  number = {12},
  pages = {1723--1732},
  issn = {1077-2626},
  doi = {10.1109/TVCG.2014.2346677},
  url = {http://ieeexplore.ieee.org/document/6875959/},
  urldate = {2025-08-31},
  langid = {english},
}

@inproceedings{duggan2011skim,
author = {Duggan, Geoffrey B. and Payne, Stephen J.},
title = {Skim reading by satisficing: evidence from eye tracking},
year = {2011},
isbn = {9781450302289},
publisher = {Association for Computing Machinery},
address = {New York, NY, USA},
url = {https://doi.org/10.1145/1978942.1979114},
doi = {10.1145/1978942.1979114},
booktitle = {Proceedings of the SIGCHI Conference on Human Factors in Computing Systems},
pages = {1141–1150},
numpages = {10},
location = {Vancouver, BC, Canada},
series = {CHI '11}
}

@inproceedings{amershiGuidelinesHumanAIInteraction2019,
  title = {Guidelines for {{Human-AI Interaction}}},
  booktitle = {Proceedings of the 2019 {{CHI Conference}} on {{Human Factors}} in {{Computing Systems}}},
  author = {Amershi, Saleema and Weld, Dan and Vorvoreanu, Mihaela and Fourney, Adam and Nushi, Besmira and Collisson, Penny and Suh, Jina and Iqbal, Shamsi and Bennett, Paul N. and Inkpen, Kori and Teevan, Jaime and Kikin-Gil, Ruth and Horvitz, Eric},
  date = {2019-05-02},
  pages = {1--13},
  publisher = {ACM},
  location = {Glasgow Scotland Uk},
  doi = {10.1145/3290605.3300233},
  url = {https://dl.acm.org/doi/10.1145/3290605.3300233},
  urldate = {2025-08-31},
  eventtitle = {{{CHI}} '19: {{CHI Conference}} on {{Human Factors}} in {{Computing Systems}}},
  isbn = {978-1-4503-5970-2},
  langid = {english},
}

@inproceedings{ehsanWhoXAIHow2024,
  title = {The {{Who}} in {{XAI}}: {{How AI Background Shapes Perceptions}} of {{AI Explanations}}},
  shorttitle = {The {{Who}} in {{XAI}}},
  booktitle = {Proceedings of the {{CHI Conference}} on {{Human Factors}} in {{Computing Systems}}},
  author = {Ehsan, Upol and Passi, Samir and Liao, Q. Vera and Chan, Larry and Lee, I-Hsiang and Muller, Michael and Riedl, Mark O},
  date = {2024-05-11},
  pages = {1--32},
  publisher = {ACM},
  location = {Honolulu HI USA},
  doi = {10.1145/3613904.3642474},
  url = {https://dl.acm.org/doi/10.1145/3613904.3642474},
  urldate = {2025-08-31},
  eventtitle = {{{CHI}} '24: {{CHI Conference}} on {{Human Factors}} in {{Computing Systems}}},
  isbn = {979-8-4007-0330-0},
  langid = {english},
}

@inproceedings{longAILiteracyFinding2023,
  title = {{{AI Literacy}}: {{Finding Common Threads}} between {{Education}}, {{Design}}, {{Policy}}, and {{Explainability}}},
  shorttitle = {{{AI Literacy}}},
  booktitle = {Extended {{Abstracts}} of the 2023 {{CHI Conference}} on {{Human Factors}} in {{Computing Systems}}},
  author = {Long, Duri and Roberts, Jessica and Magerko, Brian and Holstein, Kenneth and DiPaola, Daniella and Martin, Fred},
  date = {2023-04-19},
  pages = {1--6},
  publisher = {ACM},
  location = {Hamburg Germany},
  doi = {10.1145/3544549.3573808},
  url = {https://dl.acm.org/doi/10.1145/3544549.3573808},
  urldate = {2025-08-31},
  eventtitle = {{{CHI}} '23: {{CHI Conference}} on {{Human Factors}} in {{Computing Systems}}},
  isbn = {978-1-4503-9422-2},
  langid = {english},
}

@article{pirolliSensemakingProcessLeverage,
  title={The sensemaking process and leverage points for analyst technology as identified through cognitive task analysis},
  author={Pirolli, Peter and Card, Stuart},
  booktitle={Proceedings of international conference on intelligence analysis},
  volume={5},
  number={1},
  pages={2--4},
  year={2005},
  organization={McLean, VA, USA}
}

@article{cockburnReviewOverview+detailZooming2009,
  title = {A Review of Overview+detail, Zooming, and Focus+context Interfaces},
  author = {Cockburn, Andy and Karlson, Amy and Bederson, Benjamin B.},
  date = {2009-01-15},
  journaltitle = {ACM Computing Surveys},
  shortjournal = {ACM Comput. Surv.},
  volume = {41},
  number = {1},
  pages = {1--31},
  issn = {0360-0300, 1557-7341},
  doi = {10.1145/1456650.1456652},
  url = {https://dl.acm.org/doi/10.1145/1456650.1456652},
  urldate = {2025-09-01},
  langid = {english},
}

@inproceedings{horvitzPrinciplesMixedinitiativeUser1999,
  title = {Principles of Mixed-Initiative User Interfaces},
  booktitle = {Proceedings of the {{SIGCHI}} Conference on {{Human}} Factors in Computing Systems the {{CHI}} Is the Limit - {{CHI}} '99},
  author = {Horvitz, Eric},
  date = {1999},
  pages = {159--166},
  publisher = {ACM Press},
  location = {Pittsburgh, Pennsylvania, United States},
  doi = {10.1145/302979.303030},
  url = {http://portal.acm.org/citation.cfm?doid=302979.303030},
  urldate = {2025-09-01},
  eventtitle = {The {{SIGCHI}} Conference},
  isbn = {978-0-201-48559-2},
  langid = {english}
}

@article{hou2025thinkprune,
  title={Thinkprune: Pruning long chain-of-thought of llms via reinforcement learning},
  author={Hou, Bairu and Zhang, Yang and Ji, Jiabao and Liu, Yujian and Qian, Kaizhi and Andreas, Jacob and Chang, Shiyu},
  journal={arXiv preprint arXiv:2504.01296},
  year={2025}
}

@inproceedings{liaoQuestioningAIInforming2020,
  title = {Questioning the {{AI}}: {{Informing Design Practices}} for {{Explainable AI User Experiences}}},
  shorttitle = {Questioning the {{AI}}},
  booktitle = {Proceedings of the 2020 {{CHI Conference}} on {{Human Factors}} in {{Computing Systems}}},
  author = {Liao, Q. Vera and Gruen, Daniel and Miller, Sarah},
  date = {2020-04-21},
  eprint = {2001.02478},
  eprinttype = {arXiv},
  eprintclass = {cs},
  pages = {1--15},
  doi = {10.1145/3313831.3376590},
  url = {http://arxiv.org/abs/2001.02478},
  urldate = {2025-09-08},
  langid = {english},
}

@inproceedings{russellCostStructureSensemaking1993,
  title = {The Cost Structure of Sensemaking},
  booktitle = {Proceedings of the {{SIGCHI}} Conference on {{Human}} Factors in Computing Systems  - {{CHI}} '93},
  author = {Russell, Daniel M. and Stefik, Mark J. and Pirolli, Peter and Card, Stuart K.},
  date = {1993},
  pages = {269--276},
  publisher = {ACM Press},
  location = {Amsterdam, The Netherlands},
  doi = {10.1145/169059.169209},
  url = {http://portal.acm.org/citation.cfm?doid=169059.169209},
  urldate = {2025-09-08},
  eventtitle = {The {{SIGCHI}} Conference},
  isbn = {978-0-89791-575-5},
  langid = {english},
}

@article{mohseniMultidisciplinarySurveyFramework2021,
  title = {A {{Multidisciplinary Survey}} and {{Framework}} for {{Design}} and {{Evaluation}} of {{Explainable AI Systems}}},
  author = {Mohseni, Sina and Zarei, Niloofar and Ragan, Eric D.},
  date = {2021-12-31},
  journaltitle = {ACM Transactions on Interactive Intelligent Systems},
  shortjournal = {ACM Trans. Interact. Intell. Syst.},
  volume = {11},
  number = {3--4},
  pages = {1--45},
  issn = {2160-6455, 2160-6463},
  doi = {10.1145/3387166},
  url = {https://dl.acm.org/doi/10.1145/3387166},
  urldate = {2025-09-08},
  langid = {english},
}

@online{doshivelezRigorousScienceInterpretable2017,
  title = {Towards {{A Rigorous Science}} of {{Interpretable Machine Learning}}},
  author = {Doshi-Velez, Finale and Kim, Been},
  date = {2017-03-02},
  eprint = {1702.08608},
  eprinttype = {arXiv},
  eprintclass = {stat},
  doi = {10.48550/arXiv.1702.08608},
  url = {http://arxiv.org/abs/1702.08608},
  urldate = {2025-09-09},
  langid = {english},
  pubstate = {prepublished}
}

@inproceedings{kuleszaPrinciplesExplanatoryDebugging2015,
  title = {Principles of {{Explanatory Debugging}} to {{Personalize Interactive Machine Learning}}},
  booktitle = {Proceedings of the 20th {{International Conference}} on {{Intelligent User Interfaces}}},
  author = {Kulesza, Todd and Burnett, Margaret and Wong, Weng-Keen and Stumpf, Simone},
  date = {2015-03-18},
  pages = {126--137},
  publisher = {ACM},
  location = {Atlanta Georgia USA},
  doi = {10.1145/2678025.2701399},
  url = {https://dl.acm.org/doi/10.1145/2678025.2701399},
  urldate = {2025-09-09},
  eventtitle = {{{IUI}}'15: {{IUI}}'15 20th {{International Conference}} on {{Intelligent User Interfaces}}},
  isbn = {978-1-4503-3306-1},
  langid = {english}
}

@inproceedings{haseEvaluatingExplainableAI2020,
  title = {Evaluating {{Explainable AI}}: {{Which Algorithmic Explanations Help Users Predict Model Behavior}}?},
  shorttitle = {Evaluating {{Explainable AI}}},
  booktitle = {Proceedings of the 58th {{Annual Meeting}} of the {{Association}} for {{Computational Linguistics}}},
  author = {Hase, Peter and Bansal, Mohit},
  date = {2020},
  pages = {5540--5552},
  publisher = {Association for Computational Linguistics},
  location = {Online},
  doi = {10.18653/v1/2020.acl-main.491},
  url = {https://www.aclweb.org/anthology/2020.acl-main.491},
  urldate = {2025-09-09},
  eventtitle = {Proceedings of the 58th {{Annual Meeting}} of the {{Association}} for {{Computational Linguistics}}},
  langid = {english}
}

@article{hoffTrustAutomationIntegrating2015,
  title = {Trust in {{Automation}}: {{Integrating Empirical Evidence}} on {{Factors That Influence Trust}}},
  shorttitle = {Trust in {{Automation}}},
  author = {Hoff, Kevin Anthony and Bashir, Masooda},
  date = {2015-05},
  journaltitle = {Human Factors: The Journal of the Human Factors and Ergonomics Society},
  shortjournal = {Hum Factors},
  volume = {57},
  number = {3},
  pages = {407--434},
  issn = {0018-7208, 1547-8181},
  doi = {10.1177/0018720814547570},
  url = {https://journals.sagepub.com/doi/10.1177/0018720814547570},
  urldate = {2025-09-11},
  langid = {english},
}

@inproceedings{yuUserTrustDynamics2017,
  title = {User {{Trust Dynamics}}: {{An Investigation Driven}} by {{Differences}} in {{System Performance}}},
  shorttitle = {User {{Trust Dynamics}}},
  booktitle = {Proceedings of the 22nd {{International Conference}} on {{Intelligent User Interfaces}}},
  author = {Yu, Kun and Berkovsky, Shlomo and Taib, Ronnie and Conway, Dan and Zhou, Jianlong and Chen, Fang},
  date = {2017-03-07},
  pages = {307--317},
  publisher = {ACM},
  location = {Limassol Cyprus},
  doi = {10.1145/3025171.3025219},
  url = {https://dl.acm.org/doi/10.1145/3025171.3025219},
  urldate = {2025-09-11},
  eventtitle = {{{IUI}}'17: 22nd {{International Conference}} on {{Intelligent User Interfaces}}},
  isbn = {978-1-4503-4348-0},
  langid = {english},
}

@book{shneidermanHumanCenteredAI2022,
  title = {Human-Centered {{AI}}},
  author = {Shneiderman, Ben},
  date = {2022-01},
  eprint = {https://academic.oup.com/book/41126/book-pdf/50987951/9780192659996\_web.pdf},
  publisher = {Oxford University Press},
  doi = {10.1093/oso/9780192845290.001.0001},
  url = {https://doi.org/10.1093/oso/9780192845290.001.0001},
  isbn = {978-0-19-284529-0}
}

@online{openaiOpenAIO1System2024,
  title = {{{OpenAI}} O1 {{System Card}}},
  author = {{OpenAI} and
  Jaech, Aaron and others},
  date = {2024-12-21},
  eprint = {2412.16720},
  eprinttype = {arXiv},
  eprintclass = {cs},
  doi = {10.48550/arXiv.2412.16720},
  url = {http://arxiv.org/abs/2412.16720},
  urldate = {2025-09-02},
  langid = {english},
  pubstate = {prepublished},
}

@online{yangQwen3TechnicalReport2025,
  title = {Qwen3 {{Technical Report}}},
  author = {Yang, Anand others},
  date = {2025-05-14},
  eprint = {2505.09388},
  eprinttype = {arXiv},
  eprintclass = {cs},
  doi = {10.48550/arXiv.2505.09388},
  url = {http://arxiv.org/abs/2505.09388},
  urldate = {2025-09-02},
  langid = {english},
  pubstate = {prepublished},
}

@online{Claude3S,
  title={Claude 3.7 Sonnet System Card},
  author={Anthropic},
  date={2024},
  url={https://api.semanticscholar.org/CorpusID:276612236}
}

@article{raynerMuchReadLittle,
    author = {Keith Rayner and Elizabeth R. Schotter and Michael E. J. Masson and Mary C. Potter and Rebecca Treiman},
    title ={So Much to Read, So Little Time: How Do We Read, and Can Speed Reading Help?},
    journal = {Psychological Science in the Public Interest},
    volume = {17},
    number = {1},
    pages = {4-34},
    year = {2016},
    doi = {10.1177/1529100615623267},
    note ={PMID: 26769745},
    URL = {https://doi.org/10.1177/1529100615623267},
    eprint = {  https://doi.org/10.1177/1529100615623267}
}

@incollection{hart1988development,
  title={Development of NASA-TLX (Task Load Index): Results of empirical and theoretical research},
  author={Hart, Sandra G and Staveland, Lowell E},
  booktitle={Advances in psychology},
  volume={52},
  pages={139--183},
  year={1988},
  publisher={Elsevier}
}

@inproceedings{sauro2009seq,
author = {Sauro, Jeff and Dumas, Joseph S.},
title = {Comparison of three one-question, post-task usability questionnaires},
year = {2009},
isbn = {9781605582467},
publisher = {Association for Computing Machinery},
address = {New York, NY, USA},
url = {https://doi.org/10.1145/1518701.1518946},
doi = {10.1145/1518701.1518946},
booktitle = {Proceedings of the SIGCHI Conference on Human Factors in Computing Systems},
pages = {1599–1608},
numpages = {10},
location = {Boston, MA, USA},
series = {CHI '09}
}

@online{muennighoffS1SimpleTesttime2025a,
  title = {S1: {{Simple}} Test-Time Scaling},
  shorttitle = {S1},
  author = {Muennighoff, Niklas and Yang, Zitong and Shi, Weijia and Li, Xiang Lisa and Fei-Fei, Li and Hajishirzi, Hannaneh and Zettlemoyer, Luke and Liang, Percy and Candès, Emmanuel and Hashimoto, Tatsunori},
  date = {2025-03-01},
  eprint = {2501.19393},
  eprinttype = {arXiv},
  eprintclass = {cs},
  doi = {10.48550/arXiv.2501.19393},
  url = {http://arxiv.org/abs/2501.19393},
  urldate = {2025-09-05},
  langid = {english},
  pubstate = {prepublished},
}

@online{kumarLLMPostTrainingDeep2025,
  title = {{{LLM Post-Training}}: {{A Deep Dive}} into {{Reasoning Large Language Models}}},
  shorttitle = {{{LLM Post-Training}}},
  author = {Kumar, Komal and Ashraf, Tajamul and Thawakar, Omkar and Anwer, Rao Muhammad and Cholakkal, Hisham and Shah, Mubarak and Yang, Ming-Hsuan and Torr, Phillip H. S. and Khan, Fahad Shahbaz and Khan, Salman},
  date = {2025-03-24},
  eprint = {2502.21321},
  eprinttype = {arXiv},
  eprintclass = {cs},
  doi = {10.48550/arXiv.2502.21321},
  url = {http://arxiv.org/abs/2502.21321},
  urldate = {2025-09-07},
  langid = {english},
  pubstate = {prepublished},
}

@inproceedings{zamfirescu-pereiraWhyJohnnyCant2023a,
  title = {Why {{Johnny Can}}’t {{Prompt}}: {{How Non-AI Experts Try}} (and {{Fail}}) to {{Design LLM Prompts}}},
  shorttitle = {Why {{Johnny Can}}’t {{Prompt}}},
  booktitle = {Proceedings of the 2023 {{CHI Conference}} on {{Human Factors}} in {{Computing Systems}}},
  author = {Zamfirescu-Pereira, J.D. and Wong, Richmond Y. and Hartmann, Bjoern and Yang, Qian},
  date = {2023-04-19},
  pages = {1--21},
  publisher = {ACM},
  location = {Hamburg Germany},
  doi = {10.1145/3544548.3581388},
  url = {https://dl.acm.org/doi/10.1145/3544548.3581388},
  urldate = {2025-09-07},
  eventtitle = {{{CHI}} '23: {{CHI Conference}} on {{Human Factors}} in {{Computing Systems}}},
  isbn = {978-1-4503-9421-5},
  langid = {english},
}

@online{hammoudLastAnswerYour2025,
  title = {Beyond the {{Last Answer}}: {{Your Reasoning Trace Uncovers More}} than {{You Think}}},
  shorttitle = {Beyond the {{Last Answer}}},
  author = {Hammoud, Hasan Abed Al Kader and Itani, Hani and Ghanem, Bernard},
  date = {2025-04-29},
  eprint = {2504.20708},
  eprinttype = {arXiv},
  eprintclass = {cs},
  doi = {10.48550/arXiv.2504.20708},
  url = {http://arxiv.org/abs/2504.20708},
  urldate = {2025-09-05},
  langid = {english},
  pubstate = {prepublished},
}

@online{cuadronDangerOverthinkingExamining2025,
  title = {The {{Danger}} of {{Overthinking}}: {{Examining}} the {{Reasoning-Action Dilemma}} in {{Agentic Tasks}}},
  shorttitle = {The {{Danger}} of {{Overthinking}}},
  author = {Cuadron, Alejandro and Li, Dacheng and Ma, Wenjie and Wang, Xingyao and Wang, Yichuan and Zhuang, Siyuan and Liu, Shu and Schroeder, Luis Gaspar and Xia, Tian and Mao, Huanzhi and Thumiger, Nicholas and Desai, Aditya and Stoica, Ion and Klimovic, Ana and Neubig, Graham and Gonzalez, Joseph E.},
  date = {2025-02-12},
  eprint = {2502.08235},
  eprinttype = {arXiv},
  eprintclass = {cs},
  doi = {10.48550/arXiv.2502.08235},
  url = {http://arxiv.org/abs/2502.08235},
  urldate = {2025-09-07},
  langid = {english},
  pubstate = {prepublished},
}

@article{heerInteractiveDynamicsVisual2012,
  title = {Interactive Dynamics for Visual Analysis},
  author = {Heer, Jeffrey and Shneiderman, Ben},
  date = {2012-04},
  journaltitle = {Communications of the ACM},
  shortjournal = {Commun. ACM},
  volume = {55},
  number = {4},
  pages = {45--54},
  issn = {0001-0782, 1557-7317},
  doi = {10.1145/2133806.2133821},
  url = {https://dl.acm.org/doi/10.1145/2133806.2133821},
  urldate = {2025-09-09},
  langid = {english}
}

@article{chi1994eliciting,
  title={Eliciting self-explanations improves understanding},
  author={Chi, Michelene TH and De Leeuw, Nicholas and Chiu, Mei-Hung and LaVancher, Christian},
  journal={Cognitive science},
  volume={18},
  number={3},
  pages={439--477},
  year={1994},
  publisher={Elsevier}
}

@article{mueller2021principles,
  title={Principles of explanation in human-AI systems},
  author={Mueller, Shane T and Veinott, Elizabeth S and Hoffman, Robert R and Klein, Gary and Alam, Lamia and Mamun, Tauseef and Clancey, William J},
  journal={arXiv preprint arXiv:2102.04972},
  year={2021}
}

@inproceedings{wenEnhancingSelfExplanationStudent2025,
  title = {Enhancing {{Self-Explanation}} in {{Student Learning Through Large Language Models}}},
  booktitle = {Proceedings of the 30th {{ACM Conference}} on {{Innovation}} and {{Technology}} in {{Computer Science Education V}}. 2},
  author = {Wen, Jessica and Zavaleta Bernuy, Angela and Sibia, Naaz and Petersen, Andrew and Liut, Michael},
  date = {2025-06-13},
  pages = {762--762},
  publisher = {ACM},
  location = {Nijmegen Netherlands},
  doi = {10.1145/3724389.3730790},
  url = {https://dl.acm.org/doi/10.1145/3724389.3730790},
  urldate = {2025-09-11},
  isbn = {979-8-4007-1569-3},
  langid = {english}
}

@misc{cobbeTrainingVerifiersSolve2021,
  title = {Training {{Verifiers}} to {{Solve Math Word Problems}}},
  author = {Cobbe, Karl and Kosaraju, Vineet and Bavarian, Mohammad and Chen, Mark and Jun, Heewoo and Kaiser, Lukasz and Plappert, Matthias and Tworek, Jerry and Hilton, Jacob and Nakano, Reiichiro and Hesse, Christopher and Schulman, John},
  year = {2021},
  month = nov,
  number = {arXiv:2110.14168},
  eprint = {2110.14168},
  primaryclass = {cs},
  publisher = {arXiv},
  doi = {10.48550/arXiv.2110.14168},
  urldate = {2025-08-30},
  archiveprefix = {arXiv}
}

@misc{duaDROPReadingComprehension2019,
  title = {{{DROP}}: {{A Reading Comprehension Benchmark Requiring Discrete Reasoning Over Paragraphs}}},
  shorttitle = {{{DROP}}},
  author = {Dua, Dheeru and Wang, Yizhong and Dasigi, Pradeep and Stanovsky, Gabriel and Singh, Sameer and Gardner, Matt},
  year = {2019},
  month = apr,
  number = {arXiv:1903.00161},
  eprint = {1903.00161},
  primaryclass = {cs},
  publisher = {arXiv},
  doi = {10.48550/arXiv.1903.00161},
  urldate = {2025-08-30},
  archiveprefix = {arXiv}
}

@misc{linZebraLogicScalingLimits2025,
  title = {{{ZebraLogic}}: {{On}} the {{Scaling Limits}} of {{LLMs}} for {{Logical Reasoning}}},
  shorttitle = {{{ZebraLogic}}},
  author = {Lin, Bill Yuchen and Bras, Ronan Le and Richardson, Kyle and Sabharwal, Ashish and Poovendran, Radha and Clark, Peter and Choi, Yejin},
  year = {2025},
  month = jul,
  number = {arXiv:2502.01100},
  eprint = {2502.01100},
  primaryclass = {cs},
  publisher = {arXiv},
  doi = {10.48550/arXiv.2502.01100},
  urldate = {2025-08-30},
  archiveprefix = {arXiv}
}

@inproceedings{arendt2017matters,
  title={The “y” of it matters, even for storyline visualization},
  author={Arendt, Dustin and Pirrung, Meg},
  booktitle={2017 IEEE Conference on Visual Analytics Science and Technology (VAST)},
  pages={81--91},
  year={2017},
  organization={IEEE}
}

@article{tanahashi2012design,
  title={Design considerations for optimizing storyline visualizations},
  author={Tanahashi, Yuzuru and Ma, Kwan-Liu},
  journal={IEEE Transactions on Visualization and Computer Graphics},
  volume={18},
  number={12},
  pages={2679--2688},
  year={2012},
  publisher={IEEE}
}

@article{van2011process,
  title={Process mining and visual analytics: Breathing life into business process models},
  author={van der Aalst, Wil MP and de Leoni, Massimiliano and Ter Hofstede, AH},
  journal={BPM Center Report BPM-11-15, BPMcenter. org},
  volume={17},
  pages={699--730},
  year={2011}
}

@article{hao2006business,
  title={Business process impact visualization and anomaly detection},
  author={Hao, Ming C and Keim, Daniel A and Dayal, Umeshwar and Schneidewind, J{\"o}rn},
  journal={Information Visualization},
  volume={5},
  number={1},
  pages={15--27},
  year={2006},
  publisher={SAGE Publications Sage UK: London, England}
}

@article{ragan2015characterizing,
  title={Characterizing provenance in visualization and data analysis: an organizational framework of provenance types and purposes},
  author={Ragan, Eric D and Endert, Alex and Sanyal, Jibonananda and Chen, Jian},
  journal={IEEE transactions on visualization and computer graphics},
  volume={22},
  number={1},
  pages={31--40},
  year={2015},
  publisher={IEEE}
}

@inproceedings{xu2020survey,
  title={Survey on the analysis of user interactions and visualization provenance},
  author={Xu, Kai and Ottley, Alvitta and Walchshofer, Conny and Streit, Marc and Chang, Remco and Wenskovitch, John},
  booktitle={Computer Graphics Forum},
  volume={39},
  number={3},
  pages={757--783},
  year={2020},
  organization={Wiley Online Library}
}

@article{lee2015people,
  title={How do people make sense of unfamiliar visualizations?: A grounded model of novice's information visualization sensemaking},
  author={Lee, Sukwon and Kim, Sung-Hee and Hung, Ya-Hsin and Lam, Heidi and Kang, Youn-ah and Yi, Ji Soo},
  journal={IEEE transactions on visualization and computer graphics},
  volume={22},
  number={1},
  pages={499--508},
  year={2015},
  publisher={IEEE}
}

@inproceedings{KucherK15,
  author       = {Kostiantyn Kucher and
                  Andreas Kerren},
  title        = {Text visualization techniques: Taxonomy, visual survey, and community
                  insights},
  booktitle    = {2015 {IEEE} Pacific Visualization Symposium, PacificVis 2015, Hangzhou,
                  China, April 14-17, 2015},
  pages        = {117--121},
  year         = {2015},
  crossref     = {DBLP:conf/apvis/2015},
  url          = {https://doi.org/10.1109/PACIFICVIS.2015.7156366},
  doi          = {10.1109/PACIFICVIS.2015.7156366},
  bibsource    = {dblp computer science bibliography, https://dblp.org}
}

@proceedings{DBLP:conf/apvis/2015,
  editor       = {Shixia Liu and
                  Gerik Scheuermann and
                  Shigeo Takahashi},
  title        = {2015 {IEEE} Pacific Visualization Symposium, PacificVis 2015, Hangzhou,
                  China, April 14-17, 2015},
  publisher    = {{IEEE} Computer Society},
  year         = {2015},
  url          = {https://ieeexplore.ieee.org/xpl/conhome/7145534/proceeding},
  isbn         = {978-1-4673-6879-7},
  bibsource    = {dblp computer science bibliography, https://dblp.org}
}

@inproceedings{zou2025gistvis,
  title={GistVis: Automatic Generation of Word-scale Visualizations from Data-rich Documents},
  author={Zou, Ruishi and Tang, Yinqi and Chen, Jingzhu and Lu, Siyu and Lu, Yan and Yang, Yingfan and Ye, Chen},
  booktitle={Proceedings of the 2025 CHI Conference on Human Factors in Computing Systems},
  pages={1--18},
  year={2025}
}

@article{zhou2025landscape,
  title={Landscape of thoughts: Visualizing the reasoning process of large language models},
  author={Zhou, Zhanke and Zhu, Zhaocheng and Li, Xuan and Galkin, Mikhail and Feng, Xiao and Koyejo, Sanmi and Tang, Jian and Han, Bo},
  journal={arXiv preprint arXiv:2503.22165},
  year={2025}
}

\newpage

\appendix

\section{User Study Details}
\subsection{Post-study Questionnaire}

\begin{itemize}
    \item Overall, how useful was the raw trace of the LLM? (on a scale of 1-10, where 1 is least helpful and 10 is most helpful)
    \item Overall, how useful was the visualization of the reasoning trace? (on a scale of 1-10, where 1 is least helpful and 10 is most helpful)
    \item Briefly explain the reasoning for your ratings above. (free response)
    \item Which of the two reasoning visualization to you prefer in term of usefulness?
    \item Briefly explain you ranking. (free response)
    \item In what ways could the visualization of reasoning traces be improved? (free response)
\end{itemize}

\section{Generative AI Prompts}
Following, we provide the system prompt used to generate structured reasoning traces with \retrace. The variable \texttt{input\_data.reasoning\_content} received the stepped reasoning trace json object, see Section~\ref{}.

\begin{lstlisting}[basicstyle=\ttfamily\footnotesize]
# System Persona
You are a sophisticated reasoning analysis assistant. Your task is to meticulously analyze a sequence of Language Model (LM) reasoning steps. You will identify main phases of reasoning and then break down these phases into more granular subphases, assigning categories and summaries accordingly.

## Core Task
1.  Identify the **four main phases** of reasoning in the provided steps: "Problem Definition & Scoping," "Initial Solution & Exploration," "Iterative Refinement & Verification," and "Final Decision."
2.  For each main phase, identify one or more **contiguous subphases** based on the fine-grained actions defined in the "Subphase Categories" section.
3.  Each identified subphase must be assigned a single subcategory label (e.g., `Rephrase`, `Define_Goal`).
4.  Provide a very short, descriptive summary for each main phase.
5.  Provide a very short, descriptive summary for each identified subphase.
6.  List the 0-indexed step indices that each subphase contains.
7.  The final output must be a JSON object.

## Input Data Description (Provided separately by the user)
You will be given:
-   The original "Question" the LM was responding to.
-   A list of text strings under "Reasoning Steps," where each string is an individual reasoning step from an LM's trace. These steps are implicitly 0-indexed.

## Main Phases of Reasoning
The reasoning trace will always contain structures corresponding to these four main phases, in this order. "Iterative Refinement & Verification" may not contain any subphases if no such actions occur.

1.  **Problem Definition & Scoping**
2.  **Initial Solution & Exploration**
3.  **Iterative Refinement & Verification** (This phase is optional in terms of having content; if no relevant actions occur, its `subphases` list will be empty)
4.  **Final Decision**

## Subphase Categories (Fine-Grained Actions)
Within each main phase, identify contiguous blocks of reasoning steps that correspond to the following subphase categories. Each subphase gets *one* category label.

**Subphases for "Problem Definition & Scoping":**
* `Rephrase`: The LLM rephrases the core task or problem in its own words to ensure its comprehension. (e.g., "So, what I need to do is...", "Basically, the goal is to...")
* `Define_Goal`: The LLM clearly articulates the specific objective or what kind of answer needs to be produced. (e.g., "I need to find the total.", "The aim is to calculate...")

**Subphases for "Initial Solution & Exploration":**
* `Decomposition_&_Execution`: The LLM breaks the problem into smaller steps and immediately begins solving them, verbalizing its calculations or logical deductions. This combines planning and doing. (e.g., "First, I'll calculate X. So, if X is 5...", "Let's break this down. The first part is...")
* `First_Answer`: The LLM arrives at a complete solution or answer based on its initial line of reasoning. (e.g., "Therefore, the result is 25.", "The initial answer is...")
* `Confidence_Qualification`: The LLM briefly assesses the plausibility or correctness of its first answer, often as a prelude to more thorough checking. (e.g., "That seems right.", "Hm, let me verify that...")

**Subphases for "Iterative Refinement & Verification":**
* `Pausing_to_Rethink`: The LLM explicitly signals a stop in its current flow to reconsider, often marked by interjections like "Wait...", "Hold on...", or "Alternatively...".
* `Correction`: The LLM re-evaluates its approach by questioning assumptions, verifying specific calculations, or fixing identified errors. (e.g., "I assumed Z, but what if not?", "Let me recalculate that part.", "Ah, that calculation was wrong, it should be...")
* `Re-examine`: The LLM gets stuck re-examining an assumption or calculation it has already checked, often without making new progress. (e.g., "Let me check that again...", repeating a previous verification step).
* `Try_Alternative`: The LLM explores and develops a significantly different approach or strategy to solve the problem, leading to a new interim conclusion. (e.g., "Another way to look at this is...", "What if we try a different formula?")
* `Abandonment`: The LLM decides that a current line of reasoning is not fruitful and explicitly stops pursuing it. (e.g., "No, that won't work.", "This approach is a dead end.")

**Subphases for "Final Decision":**
* `Stating_Confidence`: The LLM expresses its degree of certainty in the correctness of its chosen final answer. (e.g., "I'm confident this is correct.", "So, I think I'm sure now...")
* `Preparing_Output`: The LLM states or formats the final answer, concluding its reasoning monologue. (e.g., "So, the final answer is...", "To summarize, the result is...")

## Structuring Logic and Instructions
1.  **Analyze All Steps**: Read the entire sequence of reasoning steps.
2.  **Identify Main Phases**: Conceptually divide the entire trace into the four main phases ("Problem Definition & Scoping," "Initial Solution & Exploration," "Iterative Refinement & Verification," "Final Decision").
3.  **Identify Subphases**: Within each main phase, identify contiguous blocks of reasoning steps that correspond to one of the defined subphase categories. A new subphase begins when the specific fine-grained action changes.
4.  **Generate IDs**: Create a unique, sequential ID for each subphase (e.g., `subphase_1`, `subphase_2`) across the entire trace.
5.  **Write Summaries**:
    *   **Main Phase Summary**: For each of the four main phases, write a very short, descriptive overall summary. This summary **must capture the essence of what the LM actually did, concluded, or defined within that specific phase for the given problem instance**. It should be a concise reflection of the phase's outcome based on the provided reasoning steps, rather than a generic restatement of the phase's name or purpose.
        *   For instance, for the 'Problem Definition & Scoping' phase in your example, a good specific summary reflecting its content would be: "The LM established the goal of creating a function to check palindrome status after removing exactly one character, clarifying ambiguities around existing palindromes and the 'exactly one removal' rule."
        *   A less helpful, generic summary for the same phase would be: "Defining the problem scope and goal."
        *   Similarly, for 'Initial Solution & Exploration' in your example, a specific summary might be: "The LM explored a two-pointer approach, considered logic for handling mismatches, and developed an initial algorithm to handle cases where the input string is already a palindrome based on character sameness and length."
    *   **Subphase Summary**: For each identified subphase, write a very short, descriptive summary detailing the specific action (e.g., "Defining problem to generate new python function", "Outlining an implementation", "Recalling knowledge about fact xyz", "Reflecting on prior assumption", "Re-checking calculation algorithm", "Deciding on solution xyz").
6.  **List Step Indices**: For each subphase, provide a list of the 0-based indices of all reasoning steps it includes.

## Important Constraints
-   Every reasoning step must be assigned to exactly one subphase.
-   All steps within a single subphase must be *consecutive* (i.e., an unbroken sequence of steps from the original reasoning trace). As a result, the `step_indices` list for any subphase must consist of sequential integers (e.g., `[2, 3, 4]` or `[5]`).
-   The sequence of main phases is fixed. Subphases must fall under their respective main phase.

## Output Format Specification (Provided separately by the user)
Your output *must be a single, valid JSON object*.

## Input Data

{{ input_data.reasoning_content | tojson }}

## Output Format

## Output Format Specification (Provided separately by the user)
Your output *must be a single, valid JSON object* with this structure:
```json
{
  "reasoning_analysis": {
    "problem_definition_and_scoping": {
      "main_phase_summary": "A very short, descriptive summary of the 'Problem Definition & Scoping' phase (e.g., Clarifying the user's goal, defining inputs, and what needs to be calculated.)",
      "subphases": [
        {
          "id": "subphase_1", // Unique sequential ID
          "subcategory": "Starting_to_Think", // Single subcategory label from the taxonomy
          "summary": "A very short summary specific to this subphase (e.g., Acknowledging the prompt.)",
          "step_indices": [0]
        },
        {
          "id": "subphase_2",
          "subcategory": "Reading_the_Question",
          "summary": "Describing the identification of key inputs.",
          "step_indices": [1]
        }
        // ... more subphases for this main phase if applicable
      ]
    },
    "initial_solution_and_exploration": {
      "main_phase_summary": "A very short, descriptive summary of the 'Initial Solution & Exploration' phase (e.g., Developing and executing an initial solution plan to arrive at a first answer.)",
      "subphases": [
        {
          "id": "subphase_X", // Continue sequential ID
          "subcategory": "Making_a_Plan",
          "summary": "Outlining the initial strategy.",
          "step_indices": [2, 3]
        }
        // ... more subphases for this main phase if applicable
      ]
    },
    "iterative_refinement_and_verification": {
      "main_phase_summary": "A very short, descriptive summary of the 'Iterative Refinement & Verification' phase (e.g., Reviewing initial findings, verifying steps, exploring alternatives, or 'No iterative refinement occurred.').",
      "subphases": [
        // This list can be empty if no such actions are present.
        // Example of a subphase within this main phase:
        // {
        //   "id": "subphase_Y",
        //   "subcategory": "Try_Alternative_(Rebloom)",
        //   "summary": "Exploring a different calculation method.",
        //   "step_indices": [4, 5]
        // },
        // {
        //   "id": "subphase_Z",
        //   "subcategory": "Reexamines_(Rumination)",
        //   "reference_subphase_id": "subphase_Y", // Optional: ID of the 'Try_Alternative_(Rebloom)' being ruminated on, if applicable
        //   "summary": "Re-checking the alternative calculation method.",
        //   "step_indices": [6]
        // }
      ]
    },
    "final_decision": {
      "main_phase_summary": "A very short, descriptive summary of the 'Final Decision' phase (e.g., Consolidating findings, confirming the solution, and preparing the output.)",
      "subphases": [
        {
          "id": "subphase_N", // Continue sequential ID
          "subcategory": "Settling_on_Solution",
          "summary": "Deciding on the definitive answer.",
          "step_indices": [7]
        }
        // ... more subphases for this main phase if applicable
      ]
    }
  }
}
```
\end{lstlisting}

\end{document}